# How to disentangle psychobiological stress reactivity and recovery: A comparison of model-based and non-compartmental analyses of cortisol concentrations

Robert Miller[1,2], Jan-Georg Wojtyniak[3], Lisa J. Weckesser[2], Nina Alexander[4], Veronika Engert[5], Thorsten Lehr[3]

[1]Department of Medical Epidemiology and Biostatistics, Karolinska Institutet, Stockholm, Sweden

[2]Institute of General Psychology, Biopsychology and Psychological Methods, TU Dresden, Dresden, Germany

[3]Department of Clinical Pharmacology, Saarland University, Saarbrücken, Germany

[4]Department of Psychology, Faculty of Human Sciences, Medical School Hamburg, Hamburg, Germany

[5]Department of Social Neuroscience, Max Planck Institute for Human Cognition, Leipzig, Germany

*Keywords:* psychosocial stress, cortisol, population pharmacokinetics, differential equation model, non-compartmental analyses, statistical power

*Author note:* Correspondence concerning this article should be addressed to Robert Miller (robert.miller@tu-dresden.de / robert.miller@ki.se), ASB, Zellescher Weg 19, 01069 Dresden, Germany. The data analyzed in this article were previously reported as stated in the methods section. We thank Jens C. Pruessner for his support and courtesy to provide the data from the Montreal sample.

*Role of the funding source:* This work was partly funded by the German Research Foundation (SFB 940/2, Project B5).

*Contributors:* RM and TL designed the present study. NCA and VE supervised the collection of the analyzed cortisol samples. RM and JW devised the functional form of and implemented the proposed population pharmacokinetic model. RM drafted a first version of the manuscript that was critically revised by LJW and TL.

## 0. Abstract

This article seeks to address the prevailing issue of how to measure specific process components of psychobiological stress responses. Particularly the change of cortisol secretion due to stress exposure has been discussed as an endophenotype of many psychosomatic health outcomes. To assess its process components, a large variety of non-compartmental parameters (i.e., composite measures of substance concentrations at different points in time) like the area under the concentration-time curve (AUC) are commonly utilized. However, a systematic evaluation and validation of these parameters based on a physiologically plausible model of cortisol secretion has not been performed so far.

Thus, a population pharmacokinetic (mixed-effects stochastic differential equation) model was developed and fitted to densely sampled salivary cortisol data of 10 males from Montreal, Canada, and sparsely sampled data of 200 mixed-sex participants from Dresden, Germany, who completed the Trier Social Stress Test (TSST). Besides the two major process components representing (1) stress-related cortisol secretion (*reactivity*) and (2) cortisol elimination (*recovery*), the model incorporates two additional, often disregarded components: (3) the secretory delay after stress onset, and (4) deviations from the projected steady-state concentration due to stress-unrelated fluctuations of cortisol secretion.

The fitted model ($R^2$ = 99%) was thereafter used to investigate the correlation structure of the four individually varying, and readily interpretable model parameters and eleven popular non-compartmental parameters. Based on these analyses, we recommend to use the minimum-maximum cortisol difference and the minimum concentration as proxy measures of reactivity and recovery, respectively. Finally, statistical power analyses of the reactivity-related sex effect illustrate the consequences of using impure non-compartmental measures of the different process components that underlie the cortisol stress response.

## 1. Introduction

In hominids two major systems mediate psychophysiological responses to acute environmental stress; the sympathetic adrenal medullary (SAM) system and the hypothalamic pituitary-adrenal (HPA) axis. While responses of the SAM system are easily inducible by most effortful situations, HPA responses require more effective stressors that are characterized by unpredictability and ego-threat (Dickerson & Kemeny, 2004, see also Koolhaas et al., 2011). Although a secretory cascade of multiple hormones accompanies such HPA responses, their most popular indicator is a transient stress-related change of cortisol concentrations. This change features a considerable portion of trait variance due to gene-environment interaction (Federenko et al., 2004, Hankin et al., 2015) and will be henceforth referred to as *cortisol stress response*. It is characterized by a phase of *reactivity* ranging from a basal pre-stress concentration to the post-stress concentration peak, and a phase of *recovery*, that follows this concentration peak until the basal concentration is reached again (see Figure 1; Kirschbaum et al. 1993, Linden et al., 1997).

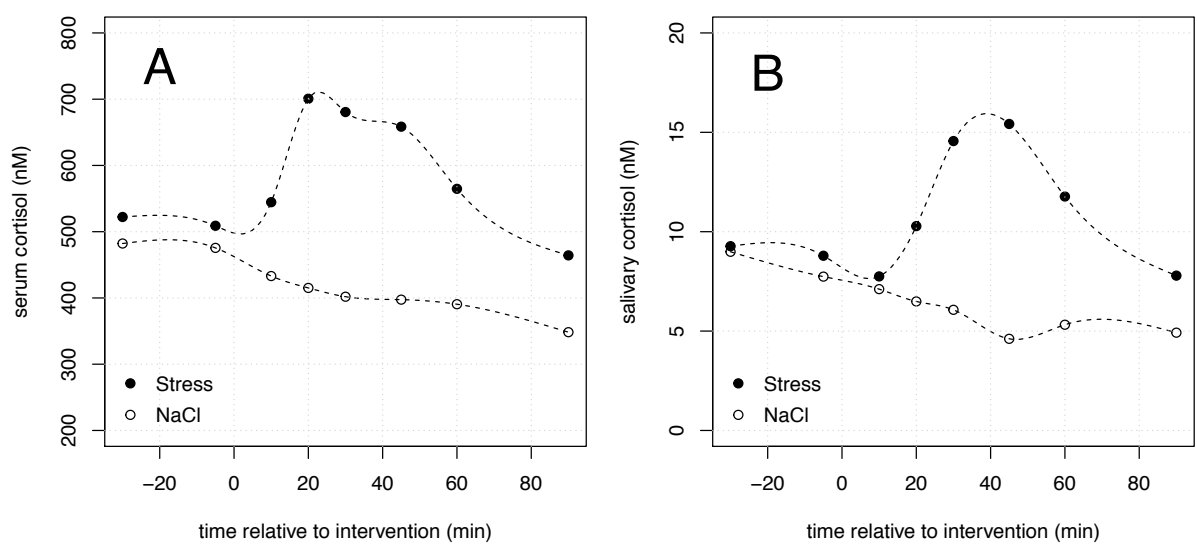

Figure 1. Change of mean cortisol concentrations in (A) blood serum, and (B) saliva in response to a stress induction protocol or bolus saline injection (NaCl). Data were obtained from 20 healthy males (age: 24.7 +- 3.3 years), who completed both interventions (Kirschbaum et al., 1993).

Apart from the superior specificity of cortisol as compared to other hormones (Koolhaas et al., 2011), another methodological advantage has promoted its popularity as the primary biomarker of psychosocial stress: Traditionally, the monitoring of cortisol concentrations has relied on the repeated sampling of blood specimens by invasive venipuncture procedures. Apart from several infrastructural disadvantages (Levine et al., 2007), however, venipuncture was found to selectively act as a potent physiological stressor (Weckesser et al., 2015) which rendered this method less practical for population-wide assessments of the cortisol stress response. In search of a more easily accessible medium, saliva specimens were found to contain lower cortisol concentrations, whose stress-related changes nonetheless mirror those observed in blood (Figure 1; Kirschbaum et al., 1993). This absolute difference between blood and salivary cortisol is attributable to the lipophilic layers of the capillary and epithelial cell tissue that seem to act as a filter for the protein-unbound fraction of steroid hormones in blood (Gröschl, 2008). Thus, salivary cortisol has become a widely-acknowledged proxy for the bioactive fraction of circulating cortisol in humans (Kirschbaum & Hellhammer, 1994, but see also Levine et al., 2007). Accordingly, it served as an ideal basis to investigate the predictive value of cortisol as an intermediate (or endo-) phenotype of stress-associated health outcomes like psychological disorders (e.g., depression), and a large variety of metabolic, and cardio- or neurovascular diseases (see Chrousos, 2009, for an overview).

## 1.1. Biometric challenges

Irrespective of any conceptual utility of the cortisol stress response for diagnostic purposes, the appropriate measurement of the underlying psychophysiological processes is still a subject of scientific debate. This is because the many time-specific cortisol concentrations $C(t)$ that are observed within $i$ individuals need to be statistically integrated into a few time-invariant process parameters $\phi_i$, which can then be related to the outcome of interest. To this end, several purpose-designed parametric models have been proposed to adequately describe the change of $C(t)$ in the specific phases of the cortisol stress response (e.g. piecewise growth curve models; Schlotz et al., 2011, Lopez-Duran, Mayer, & Abelson, 2014, or autoregressive free curve models; Miller et al., 2013). A key advantage of these models relates to their hierarchical structure, which shrinks extreme manifestations of $\phi_i$ towards their conditional mean and thereby improves the models' predictive accuracy (see Gelman et al., 2014). Nonetheless, the penetrance of these models in research on the cortisol stress response has been quite limited which is presumably promoted by a perceived low prospective gain in predictive accuracy that is weighted against the considerable implementation burden of hierarchical data modeling.

Thus, most research on the cortisol stress responses still relies on two-stage procedures that involve the estimation of $\phi_i$ using the non-compartmental pharmacokinetic analyses on the 1$^{st}$ stage (e.g. calculation of the area under the concentration-time curve, AUC; Pruessner et al., 2003, Gabrielsson & Weiner, 2012), and to subsequently predict outcomes by these $\phi_i$ on the 2$^{nd}$ stage.

Because of the very high precision of biochemical assays, these appealingly simple two-stage procedures are not considerably affected by the attenuation of potential $\phi_i$-outcome associations that would likely occur with psychometric stress markers (see Skrondal & Laake, 2001). However, high measurement precision does not completely alleviate the risk that inappropriate choices of $\phi_i$ will limit the statistical inferences that can be made:

(1) The predictive accuracy of $\phi_i$ towards an outcome (and the statistical power to detect such associations) decreases as the portion of *any* outcome-unrelated variance in $\phi_i$ increases (Hutcheon et al., 2010). Therefore, potential associations will nonetheless attenuate whenever $\phi_i$ is indicative of a mixture of unrelated processes, but only one of these processes is actually associated with the outcome of interest.

(2) The interpretability of outcome associations with $\phi_i$ is bound to the physiological validity of the underlying process model (i.e., the purity of the chosen $\phi_i$). For instance, higher AUCs do not necessarily imply an increased magnitude of the cortisol stress response, but could as well be attributed to differences in basal, stress-unrelated cortisol secretion (Balodis et al., 2010).

To evaluate the impact of these potential complications, several studies investigated the correlation structure of different $\phi_i$ using principal component analyses (Fekedulegn et al., 2007; Khoury et al., 2015). These studies found that the majority of variance in the parameters $\phi_i$ of non-compartmental pharmacokinetic analyses (i.e., 79-93%) can be attributed to two distinct biometric components of which the first probably represents the overall secretion of cortisol across time. By contrast, only the second, considerably smaller component seems to be indicative of the specific cortisol change in response to phasic events (such as an exposure to acute stress), that is not adequately reflected by many of the currently used $\phi_i$ (e.g. Khoury et al., 2015). An accurate interpretation of outcome associations with regard to the different physiological processes that are involved in the cortisol stress response further remains to be challenging, because many $\phi_i$ only serve to *describe* the apparent change of cortisol concentrations. However, they have not been validated against the parameters of a physiologically plausible model that mechanistically accounts for the underlying stress-related and stress-unrelated process components of HPA axis activity.

## 1.2. Research aims

Proceeding from the outlined biometric challenge of measuring the processes underlying the cortisol stress response, the present article seeks (A) to develop a benchmark model that is informed by pharmacokinetic theory and can therefore serve to infer the different physiological processes governing cortisol secretion in temporal proximity to acute stress exposure. Using the information provided by this model, this article further seeks (B) to accurately assess the validity of the various non-compartmental parameters $\phi_i$ that are commonly used to investigate the relation between these physiological processes and outcome variables of interest.

To achieve aim (A), the research findings and those foundations of the pharmacological compartment theory (Gabrielsson & Weiner, 2006; Bonate, 2011), that are most relevant to the modeling of cortisol secretion under basal and challenge conditions, will be summarized. Based on these foundations, a hierarchical differential equation model of the cortisol stress response is developed. In contrast to the above-mentioned growth curve models of the cortisol stress response, this novel model is inherently continuous (i.e., it accounts for the partially stochastic change of cortisol concentrations at any point in time; Voelkle et al., 2012) and adequately incorporates knowledge about the physiology of the HPA axis. Specifically, the model is supposed to yield a set of different parameters $\phi_i$ that are interpretable as the four following interindividually varying process components of acute cortisol secretion:

(1) the elimination of salivary cortisol from the organism, which determines a latent steady state of salivary cortisol that is approached in the absence of stress or other secretory pacemakers (i.e., recovery)
(2) the stress-unrelated deviation of salivary cortisol from this latent steady state at the beginning of the sampling period
(3) the magnitude of the cortisol stress response (i.e., reactivity)

    (4) the temporal delay of this response relative to the onset of stress exposure

Besides these deterministic process components, further stress-unrelated fluctuations of cortisol concentrations can occur throughout the sampling period and are accounted for by stochastic components. Thus, the developed model is not thought to be exhaustive, but only represents a simplified approximation to the most important physiological processes that operate before and after stress exposure. In consequence, the model necessarily disregards other well-known characteristics of HPA axis activity (e.g. circadian oscillations; Spiga et al., 2014) that cannot be identified by the salivary cortisol data of 210 mixed-sex individuals to which it will be subsequently fitted. The implications of these potential shortcomings are best summarized by the famous aphorism "Remember, all models are wrong; the practical question is, how wrong do they have to be to not to be useful" (Box & Draper, 1987, p. 74).

The usefulness of the developed model will be primarily shown with respect to aim (B): Proceeding from the notion that the fitted model covers the involved physiological processes (1) to (4) sufficiently well, artificial cortisol data are generated, which are representative of the study design characteristics commonly encountered in endocrine stress research. These data are then submitted to non-compartmental analyses and the correspondence of the resulting parameters to those of the data-generating model is assessed. Finally, a bootstrap is performed to demonstrate that the statistical power to detect the commonly observable sex difference in the magnitude of the cortisol stress response (i.e., males > females; Kudielka et al., 2009) is considerably reduced if $\phi_i$ is contaminated by variance from the remaining, stress-unrelated process components.

## 2. Developing a model of the cortisol stress response

Psychophysiological stress responses are obviously determined by processes comprising both, physiological and psychological characteristics. From a process modeling perspective, the psychological stress level of an individual is often thought to increase through accumulation in a psychological reservoir if environmental challenge occurs repeatedly or persists across time (e.g. Deboeck & Bergeman, 2013). In the absence of environmental challenge, by contrast, the stress level decreases across time because compensatory processes provide the individual with the constant ability to empty the reservoir. The dependency of the manifest stress level on these accumulation and dissipation processes in the basic reservoir model is schematically visualized in Figure 2.

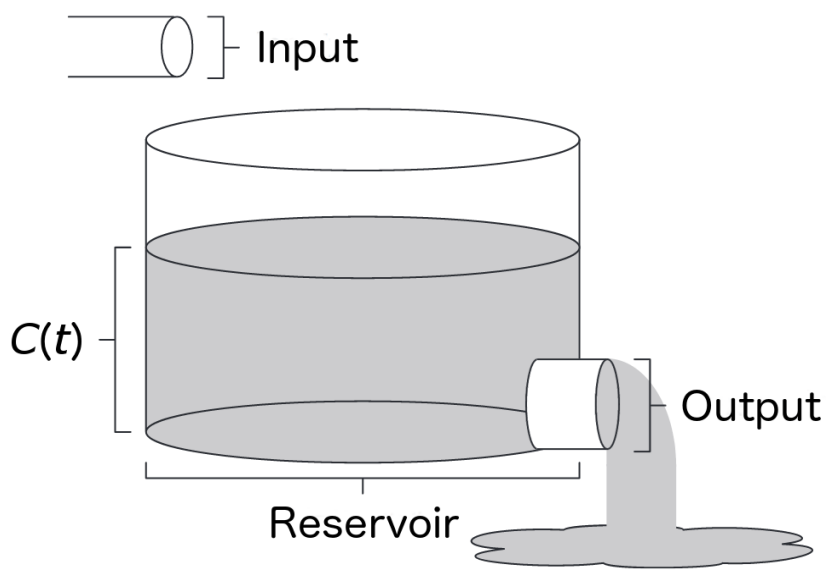

**Figure 2. Basic reservoir model of stress / cortisol at time *t* (Deboeck & Bergeman, 2013, *Psychological Methods, 18*, p.269, APA, adapted with permission).**

In pharmacological theory, such reservoirs are also known as compartments, that is, spatially separable components of a physiological system in which the concentration of a substance can be determined. Essentially, blood and saliva form such compartments[1], which is the primary reason for relying on this terminology when cortisol concentrations are modeled.

---

[1] Notably, compartments are usually characterized by a specific volume in which a substance is homogeneously distributed. The distribution volume *V* can be determined by injecting a known amount of substance into and measuring the resulting concentration in that compartment (e.g., a substance dose of 1 mmol that

Because cortisol serves as a stress biomarker, it seems reasonable to assume that its concentration changes through similar mechanisms as in the reservoir model. Thus, the accumulation of cortisol in a monitored compartment could be attributed to secretion processes (e.g., input due to stress exposure), which would be counteracted by the concurrent elimination (e.g., output due to the metabolization and excretion) of cortisol from that compartment. Such inputs and outputs can be described using ordinary differential equations (ODEs) that formalize the relative change of substance concentration *dC* per unit of time *t* (Gabrielsson & Weiner, 2006, p.105ff). The ODE of the outlined input-output model is provided below in Eq. 1, which is subject to the initial concentration *C*(0):

Eq. 1

$$\frac{dC(t)}{dt} = k_S S(t) - k_E C(t)$$

The first term of this model $k_S S(t)$ represents the secretion processes, where *S*(*t*) is an unknown function which determines the magnitude of cortisol secretion at a specific point in time, that is, *S*(*t*) simply serves as a multiplier of the secretion rate $k_S$. Accordingly, stress events could for instance unfold their phasic effects on $k_S$ through a step-function so that *S*(*t*) > 0 when stress is going, and *S*(*t*) = 0 under resting conditions. In the context of the cortisol stress response, however, such step-functions are probably too simple. Therefore, the subsections 2.1 – 2.4 will devise a physiologically plausible parametric form of *S*(*t*).

The second term of this model $k_E C(t)$ represents the elimination processes that operate on the cortisol concentration *C*(*t*) irrespective of any stress-related alteration of $k_S$. Here, $k_E$ denotes the so-called fractional turnover rate of cortisol, that can also

---

results in a blood serum concentration of 0.2 mM suggests that *V* = 1 mmol / 0.2 mmol*litre$^{-1}$ = 5 litres). Yet, *V* does not reflect a physical property, because compartments vary extremely in their capability to contain specific substances. For example, the mean effective distribution volume of the blood compartment amounts to *V* = 450 liters for bioactive cortisol (Buning et al., 2017), because its major fraction binds to carrier proteins (predominately corticosteroid-binding globulin and albumin; Lentjes & Romijn, 1999). In the present article, however, the amount of secreted cortisol is unknown and *V* therefore becomes a quantity of minor relevance.

be used to calculate the mean residence time of cortisol as $MRT = \log_e(2) / k_E$ (i.e., the average time that a cortisol molecule remains in the organism after its synthesis).

## 2.1. Tonic versus phasic cortisol secretion: The „baseline" assumption

An appealing feature of the outlined model is its capability to account for situations in which the input into the monitored compartment occurs not only in response to phasic stress events, but also due to a tonic, stress-unrelated secretion process, that results in the maintenance of a constant steady-state (or basal) cortisol concentration $C(t) = C_{SS}$ under resting conditions. From a physiological point of view, the existence of such a $C_{SS}$ seems to be very likely due to a non-zero availability of the peptide hormone ACTH, which continuously enables the entry of cholesterol into the adrenal glands where it will be subsequently converted to cortisol (Spiga et al., 2014). The ODE representation of the input-output model implies that $C(t) = C_{SS}$ when cortisol concentrations stop to change across time $dC = 0$, that is, the $C_{SS}$ is determined by the full equilibration of cortisol secretion and the elimination processes:

Eq. 2
$$\frac{dC(t)}{dt} = k_S S(t) - k_E C_{SS} = 0$$

By rearranging Eq. 2, it becomes obvious that $C_{SS}$ depends on $S(t)$, which necessarily varies across time under conditions of ongoing phasic change. Under resting conditions with $S(t) = 1$ by contrast, the time-invariant steady-state concentration can simply be calculated as $C_{SS} = k_S / k_E$, so that $k_S$ becomes interpretable as the basal secretion rate.

In order to combine the necessity of a time-varying $S(t)$, with the physiologically reasonable assumption of a $C_{SS}$, an according extension of the input-output model can be implemented by conceiving $S(t)$ as two additive subprocesses of *phasic* and *tonic* secretion $S(t) = S^*(t) + 1$. The change of cortisol in the monitored compartment is then given by

Eq. 3
$$\frac{dC(t)}{dt} = k_S(S^*(t) + 1) - k_E C(t)$$

Importantly, the time-invariant steady-state concentration $C_{SS}$ is not necessarily equal to the initial pre-stress concentration $C(0)$, that is often referred to as "baseline" cortisol. Although the assumption that $C_{SS} = C(0)$ seems to be intuitively plausible whenever the time before stress exposure can be considered as a resting period, most studies of the cortisol stress response find that $C(t)$ drops below $C(0)$ in the majority of participants (e.g. Kirschbaum et al., 1993; see Figure 1). Several explanations for such drops of $C(t)$ after stress exposure have been reported including circadian changes of the basal secretion rate (Johnson, 2007), secretory rebound (Urquhart & Li, 1969, see also Gabrielsson & Weiner, 2006, p.1019ff), and anticipatory stress (Engert et al., 2013).

Even in the absence of stress, however, phasic ACTH pulses are known to occur with a mean frequency of approximately one pulse per hour (Spiga et al., 2014). Given a mean cortisol half-life of $t_{0.5} = 40$ min in saliva (Perogamvros et al., 2011), these time intervals are not sufficient for $C(t)$ to approach a constant $C_{SS}$. This is probably also the reason why the existence of a steady state is no necessary assumption to generate plausible models of cortisol secretion (e.g. Brown et al., 2001). Nonetheless, these random perturbations of $C(t)$ highlight that $C(0)$ should be at least allowed to deviate from a $C_{SS}$ to so that the confounding of the cortisol stress response with residual stress-unrelated phasic activity of the HPA axis can be avoided.

## 2.2. Cholesterol absorption and conversion

In accordance with the previous section, tonic cortisol secretion is supposed to arise due to the availability of tonic amount of ACTH under resting conditions, whereas phasic stress events cause the phasic release of an *additional* amount of ACTH into the blood stream. Specifically, the tonic amount of ACTH enables a continuous entry of cholesterol into the adrenal glands and therefore scales the rate of cortisol secretion $k_S$ under resting conditions. By contrast, phasic ACTH pulses increase the amount of cholesterol that enters the adrenal glands, and thereby result in a transient growth of $k_S$ as determined by $S^*(t)$.

$S^*(t)$ therefore represents a time-dependent multiplier of the basal secretion rate $k_S$ and corresponds to the additional cholesterol in the unobservable compartments of the adrenal glands after stress exposure. Accordingly, an ODE model of the entry of cholesterol into the cytoplasm and its subsequent transfer and conversion into cortisol by the mitochondria of adrenal gland tissue (Spiga et al., 2014) may serve as a reasonable starting point for devising the functional form of $S^*(t)$. Such a simplified model of cholesterol change in a virtual cytoplasm compartment and a mitochondria compartment is depicted in Figure 3A. The corresponding concentration-time curves for both of these compartments are shown in Figure 3B and 3C, respectively, and illustrate the following model properties:

Initially, neither of the two compartments is assumed to contain a measurable amount of the cholesterol portion that is additionally absorbed in response to a stress-related ACTH pulse. After stress-onset at $t_0$, the cholesterol in the cytoplasm compartment (Figure 3B) then instantaneously rises to the total amount that will be converted to cortisol in response the ACTH pulse. This abrupt event has no formal physiological correspondence, but only serves to implement the gradual absorption of cholesterol into the mitochondria compartment (Figure 3C).

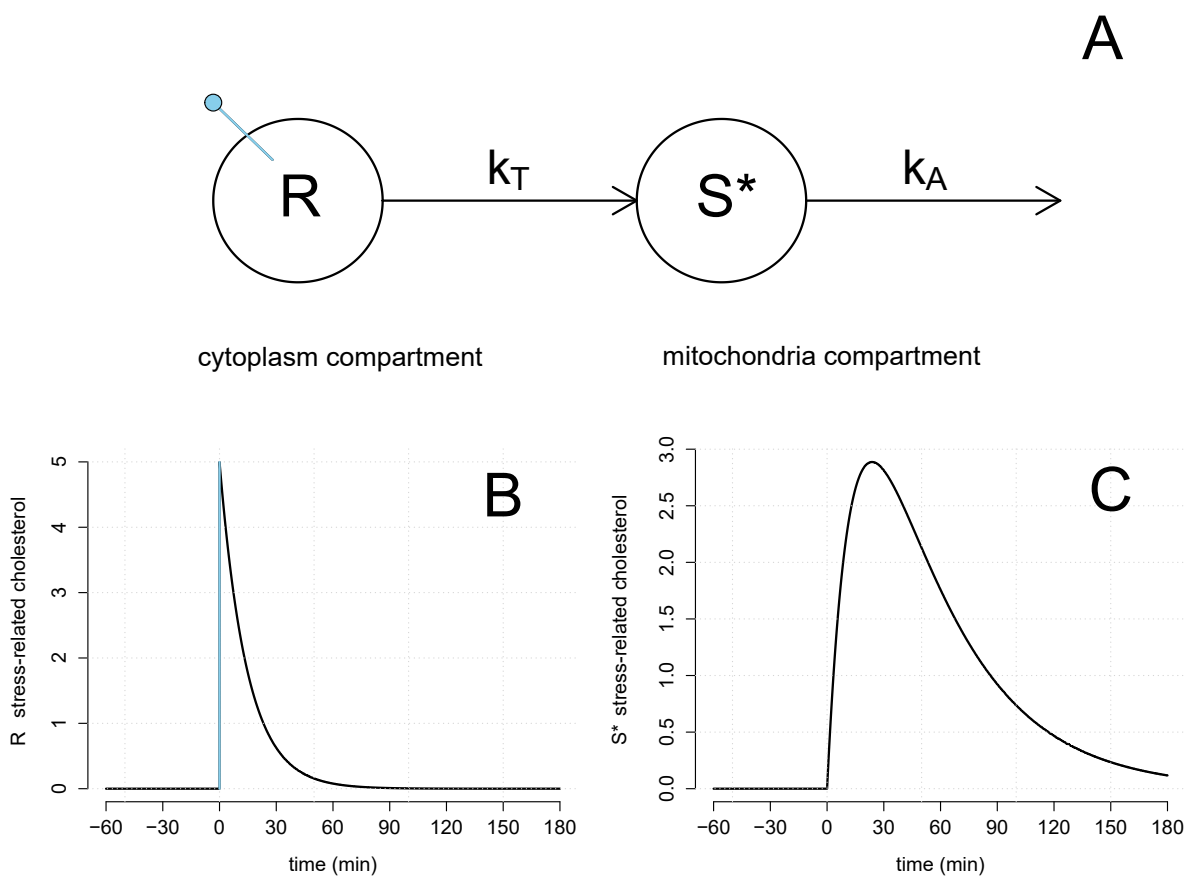

Figure 3. [A] Schematic model of cholesterol change that determines the transient increase of cortisol secretion in response to stress exposure. [B] At stress onset ($t = 0$ min), the amount of cholesterol in the cytoplasm compartment $R$ is thought to increase immediately, but will be subsequently transferred into the mitochondria compartment $S^*$ at a rate $k_T$. [C] Concurrently to its transfer, cholesterol will be converted in $S^*$ at a rate $k_A$. It is the cholesterol availability in $S^*$ at time $t$ that will finally determine the increase of cortisol secretion.

Miller et al. (in press). https://doi.org/10.1016/j.psyneuen.2017.12.019

The depicted model of $S^*(t)$ postulates that the change of cholesterol in the cytoplasm $dR$ only depends on rate of cholesterol transfer out of cytoplasm compartment $k_T$. The corresponding process is expressed by Eq. 4.

Eq. 4
$$\frac{dR(t)}{dt} = -k_T R(t)$$

By contrast, the time-dependent change of cholesterol in the mitochondria due to stress-exposure $dS^*$ is thought to depend on two different processes (i.e., cholesterol transfer and conversion) that are jointly formalized by the ODE below.

Eq. 5
$$\frac{dS^*(t)}{dt} = k_T R(t) - k_A S^*(t)$$

As can be seen, $k_T$ concurrently represents the rate of cholesterol transfer out of the cytoplasm compartment and into the mitochondria compartment. Thus, the $dS^*$ will initially correspond to $-dR$ as shown in Figure 3C. Importantly, this initial increase attenuates as less cholesterol remains in the cytoplasm compartment. Thus, the conversion process will start to dominate $dS^*$. This conversion of cholesterol into cortisol is performed at a rate parameter $k_A$, and implies that $S^*(\infty) \approx 0$ (i.e., all additional cholesterol will be converted as time goes by) if no further ACTH pulse occurs.

To enrich this basic model of phasic, stress-related cholesterol conversion by the tonic, stress-unrelated component, Eq. 5 can be generalized to Eq. 6 by substituting $S^*(t) = S(t) - 1$:

Eq. 6
$$\frac{dS(t)}{dt} = k_T R(t) - k_A S(t) + k_A$$

The system, that is formed by the Eqs. 4 and 5/6, can be analytically solved using Laplace transformations, which yields the following nonlinear equation of $S(t)$ that is determined by three easily interpretable parameters (see also Gabrielsson & Weiner, 2006, p. 31):

Eq. 7

$$S(t) = \frac{k_T R(0)}{k_T - k_A}(e^{-k_A t} - e^{-k_T t}) + 1$$

The first parameter $R(0)$ represents the relative amount of additional cholesterol that will be converted in response to the stress-related ACTH pulse. Thus, it can be conceived as the magnitude parameter of the cortisol stress response. By contrast, the shape of $S(t)$ is exclusively determined by the two rate parameters $k_T$ and $k_A$, which represent the relative change of cholesterol due to the transfer and the conversion processes, respectively.

## 2.3. Delay of stress-induced cholesterol conversion

For didactic reasons, the previous subsection relied on the idea of a cortisol conversion process that operates instantaneously on all cholesterol molecules that enter the mitochondria compartment. However, this assumption is overly simplified because it disregards that the conversion of cholesterol requires a considerable amount of time (i.e., 10 – 15 min; Spiga et al., 2014). This indirect mode of action can be modeled as a delayed transfer of cholesterol into the mitochondria compartment after ACTH exposure.

One approach to account for such a delay simply requires the estimation of an additional shift parameter μ that represents the time passing in between the onset of stress (fixed at $t_0$) and the onset of the secretory response ($t_0$ + μ). However, the utility of such change-point models is limited by (1) their physiological implausibility, and (2) difficulties in finding the best-fitting parameter set using numeric ODE solvers (Savic et al., 2007). Both shortcomings can be circumvented by the incorporation of $n$ chained transit compartments in between $R$ and $S$, which successively delay the cholesterol transfer from the cytoplasm into mitochondrial compartment (Bonate, 2011, p. 331).

Because such transit compartments are inherently unobservable, the cytoplasm compartment $R$ can simply be regarded as another transit compartment, from which cholesterol is transferred at a similar rate $k_T$ as from the remaining transit compartments. Considering the lack of knowledge about the precise kinetics of the cholesterol transfer, the cytoplasm compartment will therefore be substituted by a variable number of transit compartments (see Figure 4A). The accordingly extended ODE of cholesterol change in the mitochondria compartment (Eq. 6) is

Eq. 8
$$\frac{dS(t)}{dt} = k_T R_n(t) - k_A S(t) + k_A$$

where $R_n(t)$ denotes the relative amount of cholesterol in the last ($n^{th}$) transit compartment at time $t$, and $k_T$ represents the rate of cholesterol transfer from the last transit compartment into the mitochondria compartment.

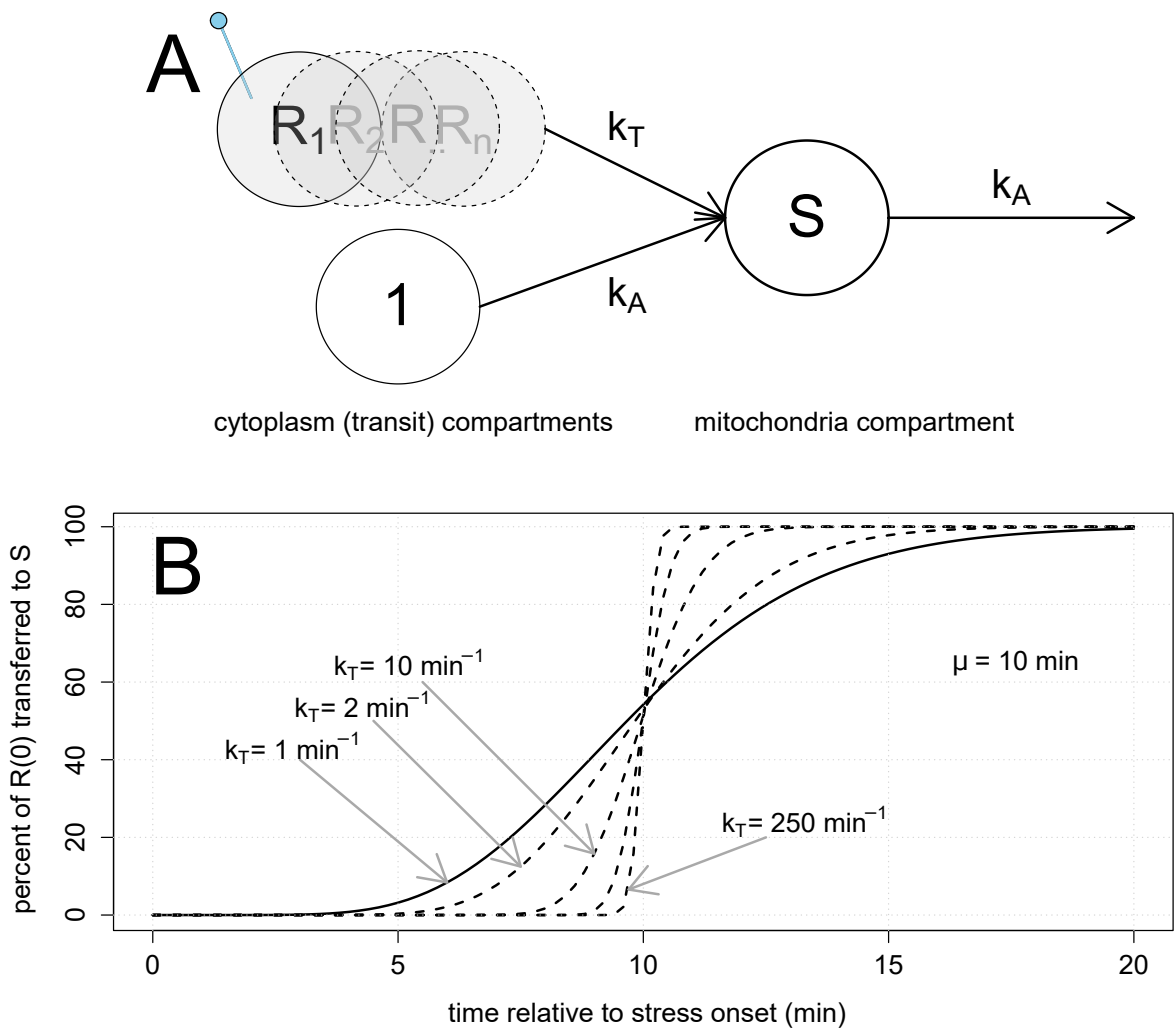

Figure 4. [A] Schematic model of $S(t)$ that accounts for a tonic, stress-unrelated cholesterol secretion (1) and the delay of phasic, stress-related cholesterol secretion using a variable number of transit compartments ($R_1 - R_n$). [B] Depending on the transfer rate ($k_T$) of cholesterol from one transit compartment to the subsequent one, the mean transit time ($\mu = [n+1]/ k_T$) of cholesterol can approximate a fixed delay time at which its complete amount will be transferred into the mitochondria compartment.

Miller et al. (in press). https://doi.org/10.1016/j.psyneuen.2017.12.019

Savic et al. (2007) reported the analytical solution for $R_n(t)$ that can be generalized to Eq. 9 by means of the Gamma function [$n! = \Gamma(n + 1)$].

Eq. 9

$$R_n(t) = R(0) \frac{k_T{}^n}{\Gamma(n+1)} t^n e^{-k_T t}$$

After plug in of Eq. 9 into Eq. 8 it becomes clear that a specific portion of $R(0)$ is transferred at each point in time as determined by the density function of the Gamma distribution $f$ with a shape parameter α = $n$+1 and a rate parameter β = $k_T$ (see appendix A):

Eq. 10

$$\frac{dS(t)}{dt} = R(0) \underbrace{\frac{k_T{}^{n+1}}{\Gamma(n+1)} t^n e^{-k_T t}}_{f(n+1,\ k_T)} - k_A S(t) + k_A$$

At this point, it should be noted, that most implementations of this model rely on its representation using a separate linear ODE for each transit compartment (Figure 4A; see also Sun & Jusko, 1998). This approach comes at the cost of reduced flexibility because it forces $n$ to be a discrete quantity. However, it concurrently decreases the time of model fitting at small $n$ due to the linearization of Eq. 10.

To facilitate parameter interpretability, the average time required for $R(0)$ to be transferred into the mitochondria compartment can either be reported as mean transit time μ = $(n+1)/k_T$ or as modal transit time $m = n / k_T$ (given $n > 0$). By contrast the dispersion of the $R(0)$ transfer with respect to time is exclusively determined by $k_T$. Accordingly, μ will become the time-invariant delay parameter of the above-mentioned change-point model, when the cholesterol transfer occurs immediately ($k_T \approx \infty$ min$^{-1}$), whereas a slower transfer entails a pronounced temporal spread of cholesterol availability in the mitochondria compartment (see Figure 4B).

### 2.4. Stochastic volatility of cholesterol conversion

So far, the developed model encompasses all specifications that are necessary to account for tonic and phasic, stress-related components of cortisol secretion, and the likely delay of cortisol secretion after stress exposure. As mentioned in subsection 2.1, however, ACTH pulses do not exclusively occur in response to stress but also due to endogenous oscillations of HPA axis activity. Although the physiological mechanisms underlying these oscillations are not completely understood, simulations suggested that they may simply emerge as a consequence of inhibition of ACTH secretion by cortisol and the secretory delay of cortisol conversion explained in subsection 2.3 (see Spiga et al., 2014, for more details). Importantly, these mechanisms will probably also operate in temporal proximity to stress and may therefore entail a misspecification bias of the parameter estimates unless countermeasures are taken.

The modeling of random deviations of the cortisol concentration from its steady-state at baseline (subsection 2.1) is such a countermeasure that can compensate for a potentially biased estimation of the magnitude of the cortisol stress response $R(0)$. The functional form of this $S(t)$, however, is also determined by the number of transit compartments $n$ and the transfer rate $k_T$, which could in principle fit any endogenous (i.e., stress-unrelated) ACTH pulse in the post-stress period that will then be mistaken for a stress-related pulse.

A parsimonious means to deal with such misspecification issues relies on the extension of the complete secretion model $dS$ by a Wiener diffusion process $dW/dt \sim N(0, \omega)$, which enables the occurrence of inert stochastic perturbations from the trajectory predicted by the Eq. 10 (e.g., Voelkle et al., 2012). The corresponding stochastic differential equation (SDE) of such an extended model is:

Eq. 11

$$\frac{dS(t)}{dt} = \underbrace{R(0)f(n+1, k_T) - k_A S(t)}_{phasic} + \underbrace{k_A}_{tonic} + \underbrace{dW/dt}_{diffusion}$$

## 2.5. Summary of the developed pharmacokinetic model

In the previous subsections, a pharmacokinetic model of the cortisol stress response was developed. This model was designed to provide parameters that have distinct physiological meanings, thereby enabling to infer the processes that govern cortisol secretion and elimination in temporal proximity to transient stress exposure.

The full model is expressed by the following ODE, which comprises the - from a physiological point of view - most relevant parameters to accurately describe the continuous change of bioavailable cortisol across time:

Eq. 12

$$\frac{dC(t)}{dt} = \underbrace{k_S\, S(t \mid R(0), \mu, k_T, k_A, \omega)}_{secretion} - \underbrace{k_E C(t)}_{elimination}$$

where $S(t)$ is determined by the SDE provided in Eq. 11. Notably, the model is further subjected to an initial condition parameter $C(0)$, that is, the cortisol concentration at baseline, which may deviate from its steady-state concentration $C_{SS} = k_S / k_E$. In sum, the model is therefore comprised of 7 structural parameters (+1 optional diffusion parameter). Table 1 lists the interpretations of these model parameters.

Table 1. Model parameters and their respective interpretations.

| Parameter | Interpretation |
| --- | --- |
| $R(0)$ | Magnitude of the cortisol stress response (relative to basal secretion) |
| $C(0)$ | Initial relative amount of cortisol (baseline cortisol) |
| $C_{SS}$ | Projected relative amount of cortisol at steady state |
| μ | Average delay for a cortisol molecule to be secreted (mean transit time) |
| $k_T$ | Rate of cholesterol transfer in response to stress exposure (per unit of time) |
| $k_A$ | Rate of cholesterol conversion (per unit of time) |
| $k_S$ | Rate of cortisol secretion at steady state (basal secretion) |
| $k_E$ | Fractional turnover rate of cortisol (per unit of time) |
| ω | Magnitude of the stochasticity in cortisol secretion (relative to basal secretion) |

*Note.* $C_{SS}$ is not directly estimable, but completely determined by the ratio of the parameters $k_S / k_E$.

In the following sections, all of these parameters will be estimated from the salivary cortisol data of 210 individuals using mixed-effects representations of the developed model. Accordingly, the between-subject variability in the most important model parameters will be quantified. The fitted model will thereafter be used to generate a large set of artificial cortisol stress responses from which the most popular non-compartmental parameters will be calculated. Based on these simulations, we will finally assess the process purity of the different non-compartmental parameters.

Miller et al. (in press). https://doi.org/10.1016/j.psyneuen.2017.12.019

## 3. Methods

### 3.1. Sample

The cortisol data that were modeled in this article are comprised of two different participant samples (total $N$ = 210) that have been investigated by Engert et al. (2011) and Alexander et al. (2014). Both studies were approved by the local Research Ethics Boards and relied on the same stress induction protocol (see subsection 3.2). The accompanying cortisol monitoring procedures only differed with respect to the frequency of saliva sampling (18 samples and 7 samples per participant, respectively).

By posting ads at McGill University in Montreal, Canada, Engert et al. (2011) recruited 50 male participants between 18 and 30 years of age who did not report a regular use of recreational drug (cannabis within the past two months, any other recreational drug within the past year) or habitual smoking (more than five cigarettes per week). Moreover, participants reporting chronic illness (including current psychosomatic disorders) or taking medication that may influence HPA axis activity were also excluded. By visual inspection 10 participants, who displayed prominent cortisol stress response (cf., Miller et al., 2013), were selected from this sample to evaluate the goodness of fit of the structural model component.

In contrast, Alexander et al. (2014) recruited 200 mixed-sex participants (50% male) between 18 and 30 years of age with a broader educational background in Dresden, Germany. Exclusion criteria were current or past mental and/or physical diseases, medication intake (for example, psychotropic drugs, substances known to influence HPA-axis activity), pregnancy, an irregular menstrual cycle and a body mass index (BMI) <17 or >30 kg/m$^2$. Considering the substantially larger sample size, it is primarily the data of these participants that informed the population variability of the model parameters.

### 3.2. Procedure

All participants were exposed to the Trier Social Stress Test (TSST; Kirschbaum et al., 1993), which is the most widely used psychological protocol to induce cortisol stress responses in laboratory settings (Goodman et al., 2017). The TSST is a social evaluative and mentally challenging test protocol that takes about 10 min for completion and yields the most robust HPA axis activations when compared to other laboratory stressors (Dickerson & Kemeny, 2004). Specifically, the protocol comprises two five-minute-phases during which the participants have to complete a 5–minute mock job interview and a 5-minute mental arithmetic task in front of an evaluating committee. To control for the exposure to food, stress, and physical exercise before starting the TSST, participants had a little snack upon arrival at the laboratory after which they rested for approx. 60 min (without eating or drinking anything but water). Since the acrophase of circadian cortisol secretion occurs proximate to awakening in humans (Stalder et al., 2016), the outlined procedure was implemented in between 1pm and 5pm, when circadian change hardly impacts on basal cortisol secretion.

### 3.3. Specimen collection and biochemical analysis

In the Montreal study, saliva specimens were collected in 10-min intervals before TSST onset (at -20, -10, and 0 min), in 2-min intervals during the TSST and the following 12 min (at +2, +4, +6, +8, +10, +14, +16, +18, +20, and +22 min, with the exception of the +12 min sample) and in 10-min intervals thereafter (at +30, +40, +50, +60, and +70 min). The +12 min specimen was skipped due to a lack of time for proper sampling when bringing participants back to their resting rooms after the TSST was completed. By contrast, the Dresden study employed a sparser sampling schedule, which yielded one specimen before TSST onset (-5 min) and six specimens after TSST onset (+11, +20, +30, +40, +55, and +70 min).

Specimens were collected using Salivette® devices (Gröschl et al., 2008) and stored at 20°C until biochemical analysis using either a time-resolved fluorescence immunoassay (DELFIA, Dressendörfer et al., 1992) for the Montreal study, or a chemiluminescence immunoassay (CLIA, IBL International) for the Dresden study. Although the IBL CLIA yields approximately 80% larger cortisol concentrations as compared to the DELFIA, the good relative correspondence between both assays and a mass spectrometric reference method has been previously demonstrated (Miller et al., 2012). All cortisol concentrations are reported in nmol/l (nM).

### 3.4. Statistical analysis

The pharmacokinetic model of the cortisol stress response that was developed in section 2 was implemented using the PSM package (Klim et al., 2009) and R 3.3.2 statistical software (R Core Team, 2016). Specifically, a nonlinear mixed-effects approach was employed to obtain population estimates of all model parameters while accounting for the between-subject variability (BSV) of these parameters (cf. Gelman et al., 2014). A concise introduction to the mixed-effects modeling of the population kinetics of pharmacological substances is provided by Mould and Upton (2012, 2013).

### 3.4.1. Model fitting and parameter estimation

The generic observation equation used for model fitting is provided below

Eq. 13

$$c_{ik} = C(t_k \mid \phi_i) + \varepsilon_{ik}$$

Here, $c$ represents the cortisol concentration of the $i^{th}$ individual that was observed at the $k^{th}$ sampling occasion. $C$ represents the cortisol concentration that was predicted at time $t$ relative to stress onset given the $p$-dimensional vector of model parameters $\phi_i$ (i.e., the individually varying solutions of Eq. 12). $\phi_i$ is determined by the population parameters $\theta = \{R(0), C(0), \mu, k_T, k_A, k_S, k_E, \omega\}$ (i.e., the fixed effects) and individual deviations $\eta_i$ from these population parameters (i.e., the random effects) as $\phi_i = \theta*\exp(\eta_i)$. The random effects are assumed to follow a multivariate Gaussian distribution $\eta_i \sim MVN(0, \Omega)$ where $\Omega$ is a symmetric covariance matrix. Finally, $\varepsilon_{ik} \sim N(0, \sigma)$ represents the normally distributed, additive residuals that are comprised of variance $\sigma^2$ due to measurement and misspecification error.

All model parameters $\Psi = \{\theta, \text{Diag}(\Omega)^{0.5}, \sigma\}$ were estimated by maximization of the models linearized likelihood function $LL(\Psi \mid c, t)$ (Wang et al., 2007). To this end, $C$ was determined using a numerical differential equation solver (Soetaert et al., 2010) that was coupled to a Fortran-coded Kalman filter (see Donnet & Samson, 2013, for an overview of different methods to estimate mixed-effects SDE models). The employed first-order conditional estimation (FOCE) algorithm involved two alternating optimization steps of which the first outer step served to find the most likely $\theta$, whereas the second inner step maximized the posterior probability of $\eta_i$ given $\Psi$.

Notably, this analysis pipeline was previously validated against the proprietary software NONMEM (Klim & Mortensen, 2008).

Robust standard errors and the 95% confidence intervals of all parameters were estimated by winsorized case bootstrapping (Yafune & Ishiguro, 1999; Ette & Onyiah, 2002)[2]. As each of the 250 employed bootstrap replicates required a fitting time of ~1.5 hours, the procedure was parallelized across 50 cores of the Linux server cluster of the Department of Medical Epidemiology and Biostatistics at Karolinska Institutet, Stockholm.

### 3.4.2. Model specification and comparison

The modeling procedure started with the fitting of the structural model part (i.e., no BSV and stochastic volatility of the secretion process were allowed) that was enriched by two further parameters to account for the relative measurement discrepancy between both assay methods, and the well-known difference between the magnitude of the cortisol stress response $R(0)$ in males (Montreal + Dresden) and females (only Dresden). Proceeding from the resulting parameter estimates, population models were generated by successively estimating the (residual) BSV of $R(0)$, $k_E$, $k_T$, and $C(0)$. Finally, the stochastic volatility parameter ω was added. The fit of these models was compared using likelihood ratio tests and Akaike weights (Vandekerckhove et al., 2015), and further evaluated by visual predictive checks (Mould & Upton, 2013). The compiled cortisol data, and the commented syntax for the specification, fitting, and comparison of all models can be downloaded from https://osf.io/ecjy6.

---

[2] One complication with the fitting of SDE models is, that numerical optimization algorithms often fail to find the global minimum. This is because the stochastic volatility term roughens the likelihood surface and thereby inflates the fit of many different parameter sets that would otherwise not have adequately accounted for the data. In such situations, the approximate Hessian is commonly not positive definite and can therefore not be used to infer on the sampling variance of the parameters. Case bootstrapping provides a solution to this issue.

## 4. Results

### 4.1. Pharmacokinetic modeling of stress-related cortisol secretion

Population pharmacokinetic modeling of cortisol concentrations in saliva before and after stress exposure was conducted as described in the subsection 3.4, using the SDE model developed throughout section 2 of the present article. The descriptive statistics of these cortisol data from 10 male participants (Montreal sample) and the 200 mixed-sex participants (Dresden) are listed in Table 2.

Table 2. Moments and quantiles of cortisol concentrations, stratified by sample.

| Time | Mean | SD | Skewness | Min | $Q_{25\%}$ | $Q_{50\%}$ | $Q_{75\%}$ | Max |
|---|---|---|---|---|---|---|---|---|
| Montreal (N = 10) | | | | | | | | |
| -20 | 4.01 | 1.26 | 0.33 | 2.38 | 3.08 | 3.94 | 4.84 | 6.40 |
| -10 | 3.66 | 1.34 | 1.78 | 2.37 | 3.24 | 3.36 | 3.65 | 7.24 |
| 0 | 4.51 | 2.48 | 0.87 | 1.63 | 3.10 | 3.66 | 5.53 | 9.97 |
| +2 | 4.49 | 2.46 | 0.62 | 1.46 | 2.89 | 3.93 | 5.62 | 9.15 |
| +4 | 4.90 | 3.21 | 1.23 | 1.41 | 3.01 | 4.21 | 5.78 | 12.71 |
| +6 | 4.88 | 2.88 | 0.74 | 1.24 | 2.99 | 4.58 | 5.50 | 10.89 |
| +8 | 5.69 | 3.61 | 0.63 | 1.68 | 2.76 | 5.09 | 6.86 | 12.53 |
| +10 | 6.51 | 4.95 | 0.95 | 1.62 | 3.05 | 5.46 | 6.79 | 16.66 |
| +14 | 8.28 | 4.93 | 0.64 | 2.90 | 4.56 | 7.27 | 9.87 | 17.84 |
| +16 | 10.08 | 5.11 | 0.71 | 4.11 | 7.51 | 9.27 | 10.75 | 19.43 |
| +18 | 10.95 | 3.90 | 0.64 | 5.89 | 8.57 | 9.89 | 11.85 | 18.06 |
| +20 | 11.77 | 3.46 | 0.35 | 7.00 | 9.75 | 11.33 | 13.07 | 17.80 |
| +22 | 12.48 | 3.14 | 1.03 | 9.50 | 10.18 | 11.69 | 13.83 | 19.56 |
| +30 | 9.18 | 2.42 | 0.11 | 5.31 | 7.45 | 9.19 | 10.45 | 13.27 |
| +40 | 6.57 | 2.02 | -0.27 | 2.77 | 5.66 | 6.72 | 7.29 | 9.56 |
| +50 | 5.15 | 1.41 | -0.96 | 2.03 | 5.10 | 5.31 | 6.11 | 6.69 |
| +60 | 4.37 | 1.38 | -0.45 | 1.61 | 3.52 | 4.52 | 5.10 | 6.46 |
| +70 | 3.48 | 1.04 | -0.53 | 1.41 | 3.07 | 3.82 | 3.91 | 5.11 |
| Dresden (N = 200) | | | | | | | | |
| -5 | 10.26 | 5.63 | 1.72 | 2.16 | 6.24 | 8.79 | 12.65 | 42.47 |
| +11 | 15.51 | 8.32 | 1.04 | 4.20 | 9.11 | 13.39 | 19.99 | 47.34 |
| +20 | 21.88 | 11.64 | 0.70 | 4.79 | 12.35 | 19.86 | 29.59 | 59.42 |
| +30 | 21.61 | 12.67 | 0.90 | 3.90 | 11.64 | 19.31 | 28.79 | 63.33 |
| +40 | 17.82 | 10.37 | 1.09 | 3.60 | 10.26 | 15.40 | 23.40 | 56.63 |
| +55 | 13.52 | 6.91 | 0.84 | 2.82 | 8.12 | 12.62 | 17.38 | 34.92 |
| +70 | 11.19 | 5.46 | 0.95 | 2.78 | 6.94 | 10.05 | 14.30 | 30.18 |

### 4.1.1. Structural model

In a first step, the pharmacokinetic model was fitted to the data without assuming any between-subject variability (BSV) of model's parameters within the investigated population or stochastic volatility of the secretion process. In order to account for the systematic differences in salivary cortisol *a priori*, the model was additionally informed by (1) a scaling factor λ, that estimated the relative discrepancy between cortisol concentrations measured by the CLIA versus the DELFIA method (Miller et al. 2012), and (2) participant sex, which was included as the strongest magnitude predictor of the cortisol stress response (Kudielka et al., 2009)[3].

As explained in subsection 2.3, the expected delay of the stress-related increase in cortisol due to its de-novo synthesis from cholesterol can be easily modeled using transit compartments. Accordingly, different model variants comprising $0 \leq n \leq 6$ transit compartments were evaluated with regard to their capability to approximate the shape of stress-related cortisol synthesis. The Akaike weights of these model variants are visualized in Figure 5A, and illustrate that more than $n = 3$ transit compartments did not further decrease the deviance between the fitted model and the observed cortisol concentrations. Accordingly, the model yielded a maximal deviance of $-2LL = 11250.9$ when $n = 0$, and a minimal deviance of $-2LL = 11181.9$ when $n = 3$ transit compartments were included.

Figure 5B shows the mean observed salivary cortisol concentrations, and their corresponding trajectories as predicted by the best fitting ($n = 3$) model. The parameter estimates of this model and their bootstrapped 95% confidence intervals are listed in Table 3 (column A). While the magnitude of the cortisol stress response was estimated to be approximately twice as large in males ($R(0)_{male} = 55.9$) as compared to females ($R(0)_{female} = 28.1 \sim 0.5*R(0)_{male}$), the cortisol concentrations that were measured using the CLIA amounted to the $\lambda = 2.5$-fold of those measured using the DELFIA method.

---

[3] To investigate the possibility that sex differences in the cortisol stress response may not be exclusively attributable to the magnitude of stress-related cortisol secretion, the model parameters $C(0)$, $k_T$, and $k_E$ were also regressed on the participants' sex. However, neither the corresponding coefficients nor the likelihood ratio ($\chi^2_{(3)} = 0.183$, $p = 0.98$) suggested the presence of further significant sex effects.

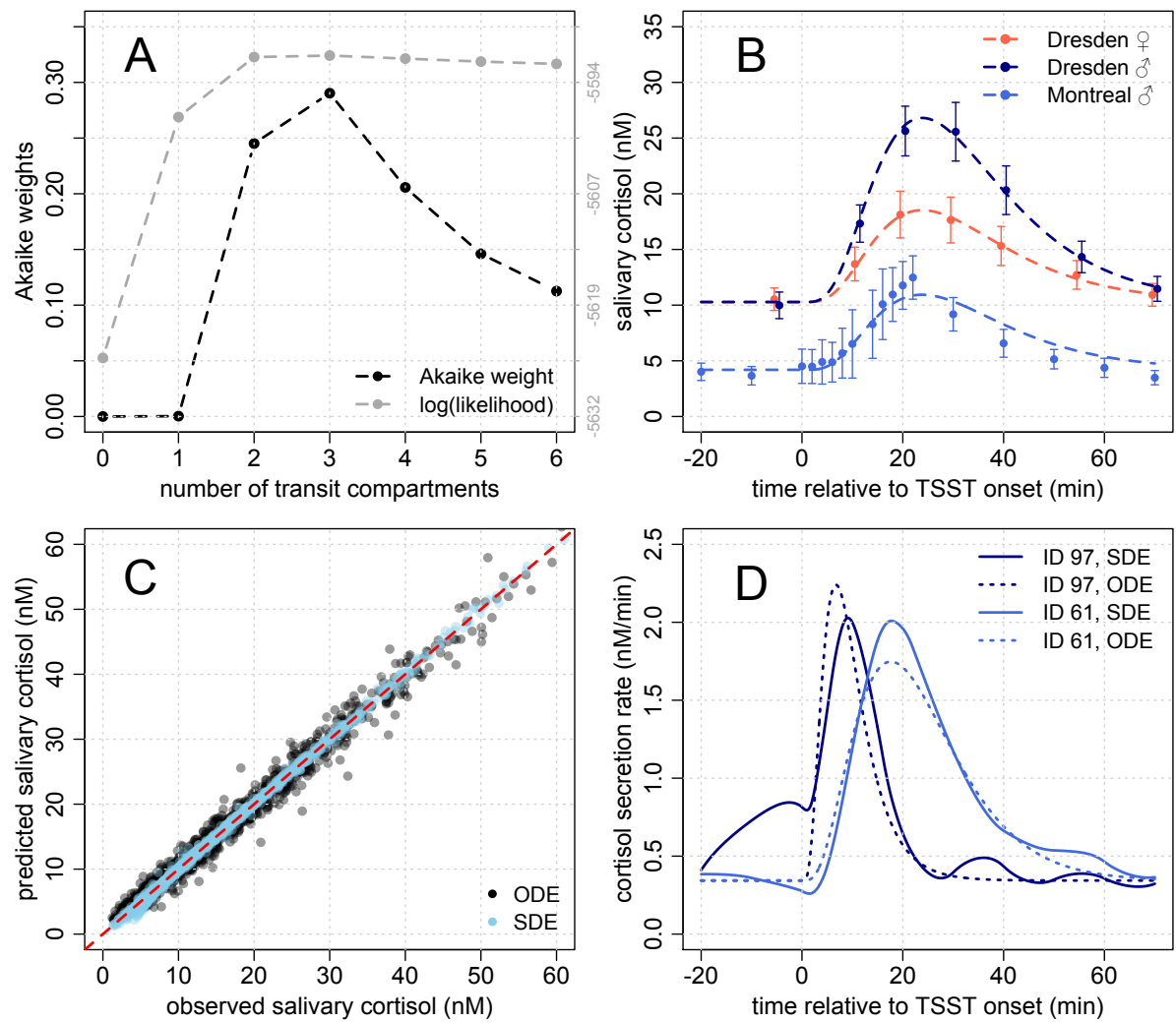

**Figure 5. (A) Comparison of structural models with different numbers of transit compartments. (B) Salivary cortisol in the Montreal and the Dresden samples. Means (± 95% CIs) of observed cortisol are indicated by points, whereas the dashed lines show the cortisol trajectories predicted by the structural model with $n = 3$ transit compartments. (C) Goodness-of-fit plot of salivary cortisol as predicted by the fully parameterized ordinary population model (ODE) and the stochastic population model (SDE). (D) Comparison of the cortisol secretion functions predicted by both models in two exemplary males from Montreal.**

All following parameter estimates are based on the DELFIA method because it was previously shown to correspond closely to mass spectrometric reference assays (Miller et al., 2012): At steady state the mean cortisol secretion amounted to $k_S$ = 0.38 nM*min$^{-1}$. As the corresponding fractional turnover rate of cortisol was estimated $k_E$ = 0.09 min$^{-1}$, the model implied a mean residence time of $MRT$ = 7.7 min and a mean steady state concentration of $C_{SS} = k_S/k_E$ = 4.20 nM. In response to stress, however, the cortisol secretion reached an average rate of 1.72 nM*min$^{-1}$ in males and 1.06 nM*min$^{-1}$ in females when the peak of the secretion function was reached at $m = n / k_T$ = 10.7 min (mean transit time: $\mu = (n+1) / k_T$ = 14.2 min).

### 4.1.2. Ordinary population model

Proceeding from the above reported results, the number of transit compartments was set to $n = 3$. In a second step, the structural model was sequentially extended by estimating the BSV of the secretory magnitude $R(0)$, the fractional turnover rate $k_E$, the transfer rate $k_T$, and finally the initial concentration $C(0) = k_S / k_E$. Table 3 lists the results of these analyses (column B-E). Notably, the precision of the fixed effects estimates increased substantially as compared to the structural model without BSV. $R(0)$ in females reduced further from 50% to 36% of the stress-related secretory magnitude in males. Moreover, $k_T$ and $k_A$ approached a similar numerical value. The goodness of fit of the fully parameterized population model is shown on Figure 5C.

Each of the added random effects accounted significantly for an incremental portion of variance in salivary cortisol and finally yielded $R^2 = 97\%$ ($-2LL = 8404.6$). Appendix B shows that this outstanding fit reflects the model's ability to selectively capture the most important process components of the stress-induced cortisol change while preserving a sufficiently low model complexity.

The BSV of $R(0)$ explained by far the largest portion of variance in salivary cortisol ($\Delta R^2 = 53\%$), with a 95% range from the 0.16 to the 6.21-fold of its sex-specific median manifestation. The BSV of $k_E$ and $k_T$ implied mean residence times of cortisol between $MRT = 5.1 - 17.2$ min, and mean transit times of cholesterol between $\mu = 7.2 - 42.8$ min, respectively. Finally, the 95% range of initial concentrations was predicted to comprise cortisol concentrations from $1.7 - 12.5$ nM (as measured by the DELFIA method).

Irrespective of the excellent fit of the full population model, the residual portion of cortisol variance suggested a coefficient of variation (CV) of 8% at 20 nM (and 16% at 10 nM). Considering the very high precision of modern immunoassays for cortisol (CVs < 6%; see Gatti et al., 2009), however, such large CVs are unlikely to be exclusively attributable to measurement error, but probably arise from the failure to completely account for non-random (temporally stable) deflections from the predicted cortisol trajectories. Detailed plots of the estimated secretion functions, the predicted trajectories, and the observed data of all 210 participants are presented in appendices C/D.

**Table 3. Pharmacokinetic models (A-F) of the cortisol stress response and their respective parameters.**

| | A Estimate | 95% CI | B Estimate | C Estimate | D Estimate | E Estimate | 95% CI | F Estimate | 95% CI |
|---|---|---|---|---|---|---|---|---|---|
| **Fixed effects (θ)** | | | | | | | | | |
| $R(0)$ | 55.92 | [44.01, 67.83] | | | | | | | |
| $R(0)_{Male}$ | | | 72.50 | 57.48 | 55.83 | 51.61 | [45.73, 57.49] | 40.90 | [33.67, 48.13] |
| $R(0)_{Female}$ | 28.05 | [20.28, 35.05] | 36.73 | 29.12 | 31.57 | 18.61 | [14.91, 22.27] | 21.56 | [19.06, 24.03] |
| $C(0) = C_{SS}$ | 4.20† | [3.37, 5.03] | 3.80† | 4.22† | 4.01† | 4.64† | [4.42, 4.85] | 4.29† | [3.88, 4.65] |
| $\mu$ | 14.21† | [12.39, 16.59] | 8.93† | 11.64† | 12.53† | 17.57† | [16.28, 19.02] | 18.28† | [15.25, 22.76] |
| $k_T$ | 0.28 | [0.24, 0.32] | 0.45 | 0.34 | 0.32 | 0.23 | [0.21, 0.25] | 0.22 | [0.18, 0.26] |
| $k_A$ | 0.09 | [0.07, 0.11] | 0.04 | 0.05 | 0.06 | 0.21 | [0.19, 0.23] | 0.21 | [0.14, 0.29] |
| $k_S$ | 0.38 | [0.28, 0.48] | 0.37 | 0.57 | 0.54 | 0.34 | [0.31, 0.38] | 0.38 | [0.27, 0.50] |
| $k_E$ | 0.09 | [0.08, 0.10] | 0.10 | 0.14 | 0.14 | 0.07 | [0.07, 0.08] | 0.09 | [0.06, 0.12] |
| $\lambda$ | 2.46 | [1.99, 3.00] | 2.16 | 2.25 | 2.46 | 1.92 | [1.85, 1.97] | 2.16 | [1.95, 2.41] |
| **Random effects (Ω)** | | | | | | | | | |
| $R(0)$ | 0† | — | 1.15 | 0.46 | 0.40 | 0.87 | [0.74, 1.00] | 0.57 | [0.33, 0.82] |
| $C(0)$ | 0† | — | 0† | 0† | 0† | 0.25 | [0.24, 0.27] | 0.18 | [0.14, 0.21] |
| $k_T$ | 0† | — | 0† | 0† | 0.21 | 0.21 | [0.18, 0.24] | 0.08 | [0.06, 0.10] |
| $k_E$ | 0† | — | 0† | 0.11 | 0.08 | 0.10 | [0.08, 0.11] | 0.10 | [0.06, 0.15] |
| **Residual variability** | | | | | | | | | |
| $\sigma^2$ | 69.36 | [56.45, 82.26] | 16.56 | 12.43 | 10.61 | 2.62 | [2.39, 2.84] | 0.64 | [0.50, 0.79] |
| $\omega$ | 0† | — | 0† | 0† | 0† | 0† | — | 0.37 | [0.34, 0.40] |
| **Model fit** | | | | | | | | | |
| $R^2$ | 0.31 | | 0.84 | 0.88 | 0.89 | 0.97 | | 0.99 | |
| LL | -5590.9 | | -4785.2 | -4651.0 | -4572.5 | -4202.3 | | -4187.1 | |
| AIC | 11197.9 | | 9588.3 | 9322.0 | 9167.1 | 8428.6 | | 8400.2 | |

*Note.* † fixed / constrained parameter. Model (A) represents the structural model. Models (B-E) estimate the between-subject variability (BSV) in some of these structural parameters. Model (F) further allows for stochastic perturbations of the cortisol trajectories.

### 4.1.3. Stochastic population model

In a final step, the portion of error-related residual variance was separated from true stochastic perturbations of the secretion process (i.e., residual autocorrelation) that were not accounted for by the ordinary population model. Table 3 (column F) lists the corresponding parameter estimates and their bootstrapped 95% confidence intervals. Due to the additional stochastic component, the model-implied residual CV reduced to 4% at 20 nM (8% at 10 nM), which conforms with the assay precision that can be expected based on previous findings (Gatti et al., 2009). This reduction of residual variance is also reflected by the goodness of fit plot in Figure 5C. Noteworthy, the stochastic volatility parameter ω accounted for a considerable portion of cortisol variance that was previously attributed to BSV in the magnitude of stress-related cortisol secretion (-34.5% in $R(0)$) and transit rates (-61.9% in $k_T$), whereas the variability in the fractional turnover rates hardly changed (+6.9% in $k_E$). Conversely, the estimates of some fixed effects were also adjusted.

The outlined effect pattern supports the idea that the stochastic population model was actually able to compensate for misspecification bias due to mechanistic simplifications in its nested ordinary population model. Proceeding from existing psychophysiological knowledge, such simplifications could relate to anticipatory stress (Engert et al., 2013) or different sources of secretory rebound (Urquhart & Li, 1969; Spiga et al., 2014) that lead to response asynchrony and the oscillatory secretion patterns after stress cessation illustrated in Figure 5D.

In support of these alleged benefits of the stochastic population model, Figure 6 further highlights its predictive accuracy as compared to the ordinary population model: While the 95% concentration range predicted by the ordinary model corresponds quite well to the time-specific distribution of cortisol in females, it substantially overestimates the dispersion of cortisol in males. By contrast, the predictions of the stochastic model correspond much better to the observed data. This is also because the ordinary model suggests a much larger sex difference in the magnitude of stress-related cortisol secretion [$R(0)_{female} = 0.36*R(0)_{male}$] as compared to the stochastic model [$R(0)_{female} = 0.53*R(0)_{male}$], which yields an effect that was much closer to the original estimate of the structural model without BSV.

Miller et al. (in press). https://doi.org/10.1016/j.psyneuen.2017.12.019

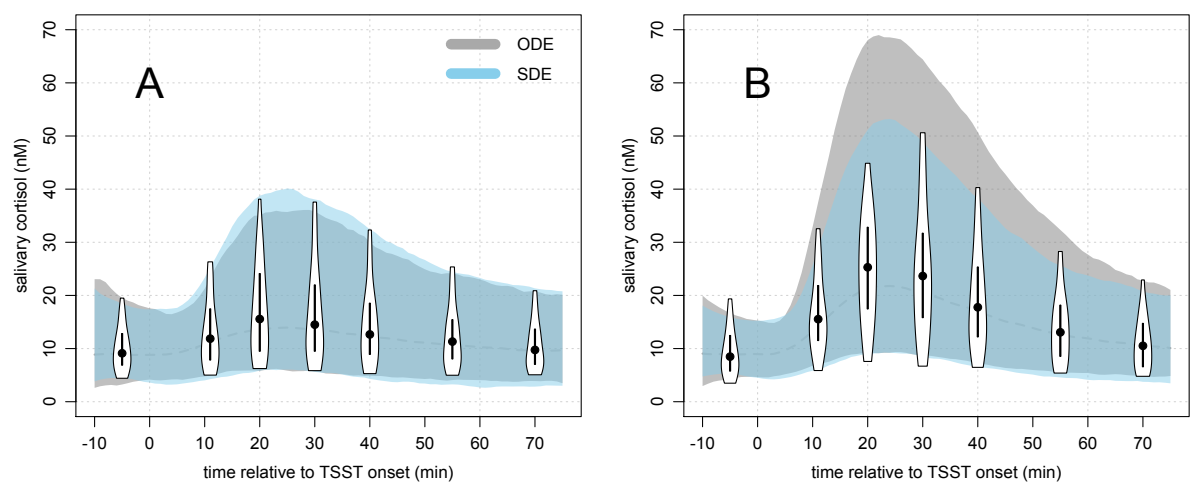

Figure 6. Visual checks of model predictions for (A) females and (B) males from Dresden. Points represent the observed median concentration, whereas the error bars encompass the IQR (i.e., the range from 25% – 75% quantile) of observed salivary cortisol. The violins and the shaded regions encompass the range from the 2.5% – 97.5% quantile of observed and model-implied salivary cortisol, respectively. ODE = ordinary differential equation, SDE = stochastic differential equation

## 4.2. Comparison of model-based and non-compartmental analyses

### 4.2.1. Correspondence between different parameters

In order to investigate if cortisol concentrations at specific points in time and their commonly encountered non-compartmental parameters (see Khoury et al., 2015) actually reflect the different process components of the cortisol stress response, a simulation study was performed. Proceeding the fitted stochastic population model (section 4.1.3), we simulated the cortisol trajectories of 10,000 virtual individuals (50% females) during the time period from -20 min to 80 min relative to TSST onset (sampling interval: 2 min). Accordingly, we obtained 50 by 10,000 artificial cortisol concentrations (DELFIA method).

The time-dependent rank correlations between these cortisol concentrations and each of the four inter-individually varying parameters of the data-generating model are shown in Figure 7A. $C(0)$, $R(0)$, and $k_E$ correlated substantially ($|\rho\text{'s}| \sim 0.7$) with salivary cortisol at -20 min, 20–30 min, and 50–80 min, respectively. By contrast, $k_T$ was hardly correlated with salivary cortisol at any point in time ($|\rho\text{'s}| < 0.3$).

Table 4 provides a complete list of all investigated non-compartmental parameters (incl. their population characteristics) that were calculated from these cortisol data. Most of these parameters were adopted from the literature review of Khoury and colleagues (2015), who claimed that the correlation structure of these parameters was primarily attributable to two distinct biometric components representing total cortisol secretion, and the stress-related secretory change, respectively. Beyond of this empirical finding, however, the data-generating model further implied the existence of two additional, meaningful variance components that reflect the delay of the cortisol stress response, and the initial deflection of cortisol levels from their steady state. As these two remaining components were insufficiently represented by the non-compartmental parameters that were investigated by Khoury et al. (2015), the cortisol concentration at the beginning of the sampling period (*Cinit*) and the time of the concentration peak (*Tmax*) were also determined (Fekedulegn et al., 2007).

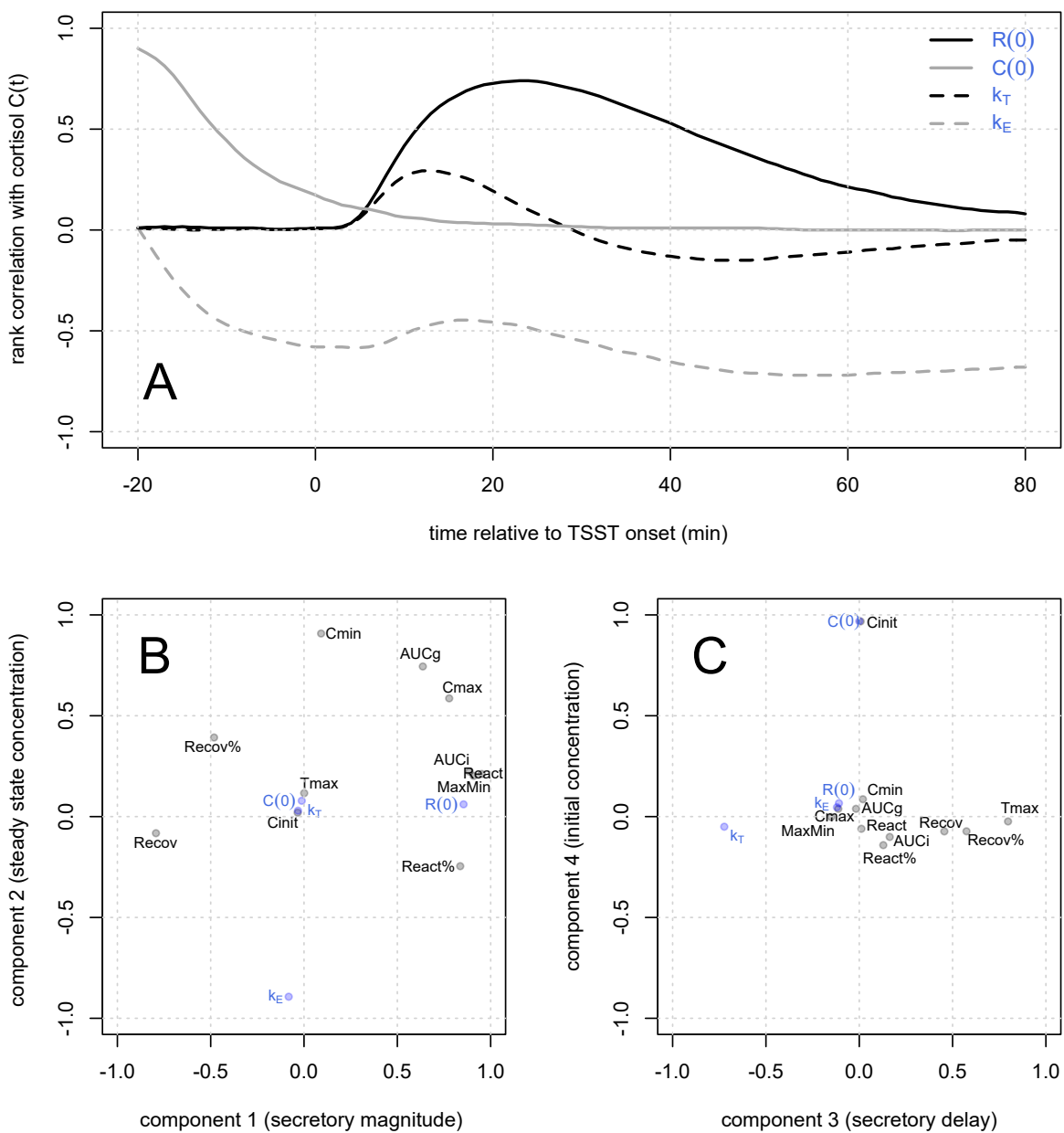

Figure 7. (A) Correlations between salivary cortisol at time *t* and the four parameters of the SDE model comprising between-subject variability. (B, C) Loadings of various parameters obtained by non-compartmental analysis (black; see Table 4) and the data-generating SDE parameters (blue) on their first four principal components.

In accordance with the outlined reasoning, a principal component analysis indicated that 83% of the correlation between all of these parameters could be explained by four orthogonal variance components with a *RMSR* = 0.06. The eigenvalues of all further components fell below Kaiser-Guttman criterion and the limit determined by parallel analyses of uncorrelated variables (Humphreys et al., 1975). After varimax rotation, the first four components were interpretable as (1) magnitude of stress-related cortisol secretion (or reactivity), (2) steady state concentration arising from the equilibration of basal cortisol secretion and elimination (or recovery), (3) secretory delay, and (4) stress-unrelated fluctuation of the initial cortisol concentration.

The loadings of each parameter on these four components are visualized in Figure 7B/C, and confirm that the non-compartmental parameters *MaxMin*, *React*, and *AUCi* (cf. Table 4) were good proxy measures of $R(0)$, whereas *Cmin* was the only suitable proxy of $k_E$. This latter finding is particularly interesting, because the time of *Cmin* occurred in 43% of all cortisol responses up to 60 min *after* stress offset, indicating that the steady state of salivary cortisol is quite often not appropriately indexed by a single "baseline" concentration like *Cinit*. Instead, *Cinit* was an exclusive proxy measure of $C(0)$ because any phase-asynchronous, stress-unrelated secretory activity had necessarily subsided when the recovery phase was reached (see also Figure 7A).

Finally, $k_T$ was best indicated by *Tmax*, although this association was sensitive to sampling frequency: Since 50% of all cortisol peaks occurred in between 20 min ≤ *Tmax* ≤ 32 min, an increase of the sampling interval to 15 min resulted in a considerable decrease of correlation $\Delta\rho(k_T, Tmax) = 0.07$. Although the $k_T$-associated parameters *Recov* and *Recov%* were less sensitive as compared to *Tmax*, they exhibited substantial cross-loadings on the 1st (or the 2nd) component which disqualified them as appropriate proxy measures. The correlation decrease of the other surrogates under conditions of lower sampling frequencies was mostly negligible except for $\Delta\rho(R(0), MaxMin) = 0.05$.

Table 4. Descriptive statistics of various parameters from non-compartmental analyses that have been compiled from 10,000 simulated cortisol stress responses (first sample: -20 min relative to TSST onset, sampling interval: 2 min), and their associations with the parameters of the data-generating population SDE model.

| Parameter | Unit | Definition | Quantiles | | | Rank correlation with | | | |
|---|---|---|---|---|---|---|---|---|---|
| | | | 25% | Median | 75% | R(0) | C(0) | $k_E$ | $k_T$ |
| Cinit | nM | $C_{-20}$ | 3.13 | 4.31 | 5.76 | 0.01 | 0.44 | -0.47 | 0.00 |
| Cmin | nM | $Min(C_t)$ | 1.55 | 2.65 | 3.96 | 0.19 | 0.09 | -0.71 | -0.02 |
| Cmax | nM | $Max(C_t)$ | 7.78 | 9.96 | 13.30 | 0.72 | 0.02 | -0.54 | 0.08 |
| Tmax | min | Time of $Max(C_t)$ | 20 | 24 | 32 | -0.11 | -0.03 | -0.21 | -0.41 |
| MaxMin | nM | $Max(C_t) - Min(C_t)$ | 5.46 | 7.71 | 9.87 | 0.78 | -0.03 | -0.27 | 0.11 |
| React% | % | $Max(C_t) / C_0$ | 180 | 244 | 359 | 0.58 | -0.13 | 0.08 | 0.05 |
| Recov% | % | $C_{60} / Max(C_t)$ | 35 | 49 | 64 | -0.42 | -0.01 | -0.34 | -0.21 |
| React | nM | $Max(C_t) - C_0$ | 3.66 | 5.70 | 8.77 | 0.77 | -0.07 | -0.28 | 0.08 |
| Recov | nM | $C_{60} - Max(C_t)$ | -7.58 | -4.87 | -3.13 | -0.72 | -0.02 | 0.10 | -0.18 |
| AUCg | nM*min | $\int C_t\,dt$ [1] | 294.40 | 389.91 | 523.08 | 0.61 | 0.03 | -0.68 | 0.02 |
| AUCi | nM*min | $\int C_t\,dt - C_0*60$ min | 52.42 | 138.41 | 249.44 | 0.69 | -0.09 | -0.30 | 0.01 |

*Note.* $C_t$ comprises all cortisol concentrations in between 0 and 60 min relative to TSST onset. [1] The integral of the concentration-time curve is approximated using the trapezoidal decomposition of the linearly interpolated concentrations (see Pruessner et al., 2003).

### 4.2.2. Statistical power for detecting parameter-outcome associations

Although the above presented analyses of simulated cortisol data suggest that the commonly used non-compartmental parameters actually reflect the different process components of the cortisol stress response to a considerable extent, they also revealed that their variance is often *not* exclusively attributable to only one process component. This may have substantial implications for the probability to statistically detect associations between covariates or outcomes and these non-compartmental parameters. Specifically, the statistical power will reduce due to regression dilution (Hutcheon et al., 2010) whenever exclusively one process component drives these associations. However, the statistical power may occasionally increase if the chosen, impure non-compartmental parameter coincidentally taps into an association that is shared by several process components. Thus, non-compartmental parameters are subjected to the same advantages and disadvantages as any composite endpoint (Ferreira-Gonzáles et al., 2007).

To illustrate the more likely case of regression dilution, further analyses were performed to determine the statistical power of various non-compartmental parameters to detect the above reported sex effect on the magnitude of the cortisol stress response (subsections 4.1). Proceeding from the results of the simulation study in subsection 4.2.1, the parameters *Cmax*, *MaxMin*, *AUCi*, and *AUCg* were assumed to be similarly indicative of the magnitude of the cortisol stress response. However, they substantially differed with respect to their cross-loadings on the remaining process components, that is, *MaxMin* and *AUCi* were found to incorporate substantially less variance from the other process components as compared to *Cmax* and *AUCg*. Accordingly, *MaxMin* and *AUCi* were hypothesized to yield a larger statistical power when Spearman rank correlation tests were used to infer the association between the participants' sex and the magnitude of the cortisol stress response. By contrast, non-compartmental parameters that primarily indicated the other three process components (*Cmin*, *Cinit*, *Tmax*) were hypothesized to fail in detecting this sex effect.

To avoid any possible confounding due to misspecification bias, the statistical power of all investigated non-compartmental parameters was determined by drawing 100,000 bootstrap replicates per scenario from the Dresden sample. Figure 8 visualizes the results of these analyses. Among all investigated non-compartmental parameters indicative of the secretory magnitude, *MaxMin* yielded the largest power

Miller et al. (in press). https://doi.org/10.1016/j.psyneuen.2017.12.019

($n_{male} = n_{female} > 23$ to achieve 80%), whereas *AUCg* yielded the lowest power ($n_{male} = n_{female} > 74$ to achieve 80%). The statistical power of *Cmax* and *AUCi* varied in between these two extremes, but surprisingly *Cmax* yielded a slightly larger power as the *AUCi*, although the latter was hypothesized to incorporate less variance from the other process components. Less surprising, the model-based *R*(0) estimates exceeded the power of all investigated non-compartmental parameter.

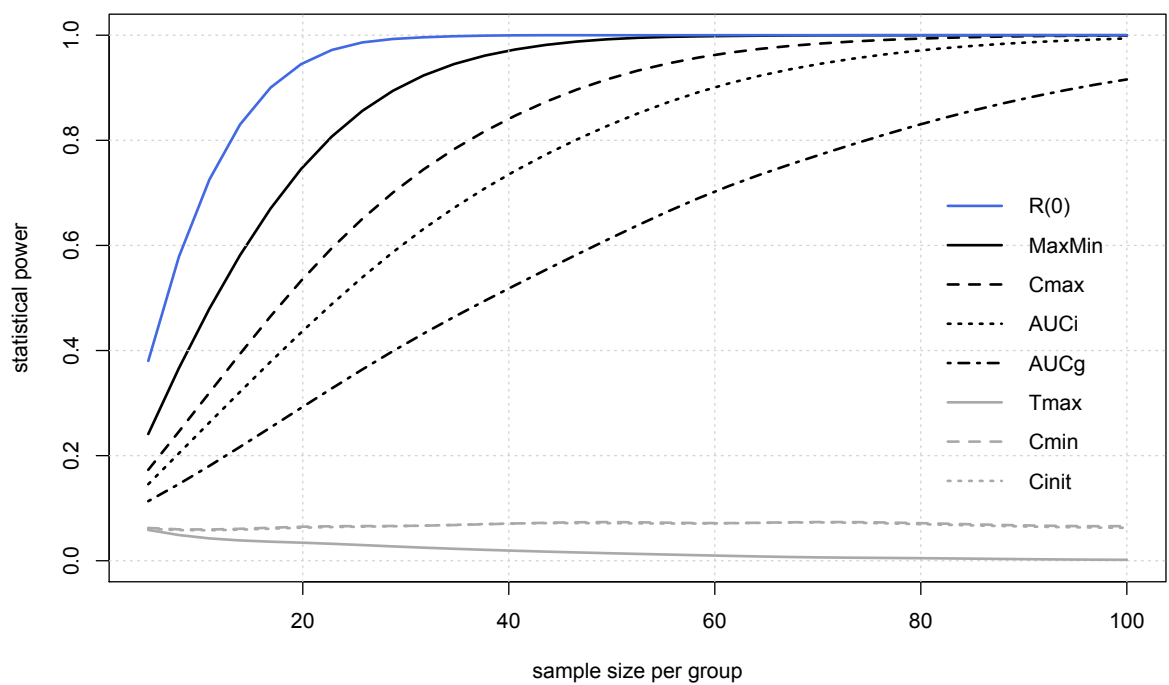

**Figure 8. Sample-size dependent power to detect the specific association (Spearman rank correlation with $p \leq 5\%$) between participant sex and magnitude of cortisol secretion using different proxy measures of cortisol secretion. The blue line represents the *R*(0) parameter of the developed pharmacokinetic model. Black lines indicate four non-compartmental parameters (*MaxMin, Cmax, AUCi, AUCg*) that primary loaded on the "secretory magnitude" component, but varied in their cross-loadings on the other components. Grey lines indicate the three non-compartmental parameters that were most indicative of the remaining three components (*Tmax*: secretory delay, *Cmin*: steady state concentration, *Cinit*: initial concentration / baseline cortisol).**

## 5. Discussion

The present article sought to address the question of how to adequately infer the different process components that govern cortisol secretion in response to psychosocial stress. While various parameters from non-compartmental analyses (e.g., change scores or the variants of the *AUC*; Khoury et al., 2015) are commonly used as proxies of these components, their specific validity to this end has not been systematically investigated, yet. Thus, a physiologically informed pharmacokinetic model was developed. This model was fitted to the salivary cortisol data of 210 mixed-sex participants who completed the TSST (Engert et al., 2011, Alexander et al., 2014), and remarkably explained up to 99% variance in all measured cortisol concentrations. In contrast the previously proposed, hierarchical growth curve models of the cortisol stress response (e.g., Schlotz et al., 2011, Lopez-Duran et al., 2014), this novel model was further designed to minimize the misspecification bias in the empirical Bayes estimates of its four interindividually varying parameters, which could therefore be used as a criterion to assess the validity of the commonly used parameters from non-compartmental analyses.

### 5.1. Physiological validity of the model and potential extensions

The minimization of misspecification bias was achieved through the estimation of stochastic perturbations in the cortisol secretion function by means of a Kalman filter that compensated for the potential mechanistic simplifications of the model. Such simplifications can for instance arise from the existence of higher order elimination kinetics. In this regard, the herein assumed first-order elimination kinetics of salivary cortisol relied on the consistent findings of other studies (e.g. Tunn et al., 1990, Perogamvros et al., 2011), whereas only the elimination of total (i.e., protein-bound + bioactive) cortisol in blood was previously shown to require second-order elimination kinetics (Kraan et al., 1997, Czock et al., 2005). In agreement with this assumption, the developed pharmacokinetic model accounted excellently for the change of salivary cortisol during the recovery phase of the Dresden sample. Notably, however, the 10 males from Montreal, who were subjected to a high-frequency sampling schedule, showed a tendency to display lower cortisol in the recovery phase than predicted by the structural part of model (Figure 5B). Although this finding may simply relate to a non-representative parameter configuration in these 10 males, it could also indicate another source of mechanistic simplification, for which no previously published data were available:

The secretion kinetics of the model assumed that phasic cortisol secretion during the reactivity phase can be mapped onto the density function of the right-skewed Gamma distribution. Thus, any stress-related cortisol secretion was hypothesized to quickly accelerate after ACTH exposure but to cease more slowly after the modal transit time had elapsed. However, the pronounced concentration peaks in the Montreal sample might be better explainable by a more flexible, left-skewed function (e.g. the density of the generalized Gamma distribution; see appendix A) so that the cortisol secretion could increase slowly after ACTH exposure, but quickly cease as soon as a certain concentration threshold is finally reached. Irrespective of the Kalman filters capability to account for this potential misspecification, its practical impact is probably negligible, because the precise kinetics of salivary cortisol during the reactivity phase can hardly be determined from the sparse sampling schedules that are commonly used in endocrine stress research. By contrast, other refinements of the structural model part (e.g. the circadian changes in the steady state concentration of salivary cortisol, Johnson, 2007; or feedback on the secretion rate of ACTH by cortisol using pharmacodynamic dose-response functions, Spiga et al., 2014) may be more worthwhile to consider, if the residual stochasticity in stress-related cortisol secretion should be mechanistically explained.

To this end, however, the linearity of the differential equations constituting the model must be abandoned, which would tremendously increase the computational costs of model fitting using the implementation chosen in this article. Thus, inclined researchers may consider to rely on other analysis pipelines, such as WinBUGS (Lunn et al., 2002), or the combination of the R package PMXStan (Xiong et al., 2015) with Stan statistical software (Gelman et al., 2016) which enable the easy implementation of fully Bayesian inference of the ODE/SDE model parameters (cf. Donnet & Samson, 2013).

### 5.2. Process components and non-compartmental parameters

The utility of the model was shown with regard to the differential correspondence between its four interindividually varying parameters and eleven common parameters from non-compartmental analyses. Associations between these parameters and covariates or outcome variables are often interpreted with respect to the process components that underlie the cortisol stress response. However, such inferences rely on the assumption that these parameters are process-pure, that is, they incorporate variance that can be exclusively attributed to the respective component.

Proceeding from the artificial cortisol data that were simulated using the proposed model, the magnitude parameter of the stress response $R(0)$ (i.e., the "reactivity" component) was best reflected by the non-compartmental parameters *MaxMin* and *React*, whereas the delay of the stress response as determined by $k_T$, the initial cortisol concentration $C(0)$, and the steady state concentration as determined by $k_E$ (i.e., the "recovery" component), were best reflected by the non-compartmental parameters *Tmax*, *Cinit*, and *Cmin*, respectively. Intriguingly, the good correspondence of $k_E$ and *Cmin* suggests that the assessment of individual differences in stress recovery does not necessarily require an active (and costly) stimulation of HPA axis activity that is followed by a long waiting period until cortisol concentrations have returned to their steady state. Instead, the cortisol levels under resting conditions (i.e., in the absence of any ultradian secretory activity) could probably also serve as a counter-intuitive proxy of stress recovery.

Although the popular non-compartmental parameter *AUCi* (Pruessner et al., 2003, Stalder et al., 2016) was initially also found to predominately reflect $R(0)$, subsequent power analyses regarding the detection of the well-known association between participant sex and the cortisol stress response (see Kudielka et al., 2009) raised some doubts about this assumption. Obviously, the *AUCi* was also comprised of variance from the other process components that were not related to $R(0)$. Thus, any association with the *AUCi* should be interpreted with caution.

A post-hoc explanation for this finding relates to the slightly different ways of how the *AUCi* was calculated from the real cortisol data of Dresden sample as compared to the artificial data that was obtained from the model-based simulations. While the simulated data allowed to calculate the *AUCi* using "baseline" cortisol concentrations at the onset of stress exposure, empiric studies commonly lack such a sampling occasion because their participants are already busy with stress anticipation at this point in time. Instead, sampling occasions prior to stress exposure are used to determine the "baseline" concentration (Balodis et al., 2010). Due to stress-unrelated ultradian activity of the HPA axis, the model predicts that such baseline data are more likely to be contaminated by stress-unrelated processes. Accordingly, it should be emphasized that any 2[nd] stage regression of the *AUCi* on a covariate of interest (e.g. participant sex) relies on the rigid and often questionable assumption that a cortisol concentration at "baseline" *Cinit* will impact in a set way on all cortisol concentrations that are subsequently sampled after stress onset:

Eq. 14
$$AUCi = AUCg - \Delta t * Cinit = \beta_0 + \beta_1 * Covariate + \varepsilon$$

which is equivalent to Eq. 15, where the set impact of "baseline" cortisol is modeled as a regression offset $\Delta t * Cinit$:

Eq. 15
$$AUCg = \beta_0 + \beta_1 * Covariate + \Delta t * Cinit + \varepsilon$$

Thus, it turns out that the assumption of a fixed "baseline" impact could be easily alleviated by adding *Cinit* as a second covariate to any regression model that investigates the association between *R*(0) as indexed the *AUCg* (which would otherwise also be contaminated by stress-unrelated secretion processes; see Figure 7B) and the primary covariate of interest:

Eq. 16
$$AUCg = \beta_0 + \beta_1 * Covariate + \beta_2 * Cinit + \varepsilon$$

Using this model specification for 2nd stage analyses of *R*(0) will likely yield a larger statistical power as compared to simple regressions on the *AUCi* (Eq. 14), and could therefore prove to be more useful for explaining additional BSV in the magnitude of the cortisol stress response.

## 6. Conclusion and Recommendations

The present article presented a physiologically plausible, differential equation model of the cortisol stress response that was used to reliably infer the four major process parameters determining the interindividually variable change of salivary cortisol due to stress exposure (i.e., secretory magnitude / *reactivity*, elimination / *recovery*, secretory delay, and stress-unrelated fluctuations). The model fitted the data exceptionally well ($R^2$ = 97-99%) and can be readily implemented using the R script that is provided as supplementary material to this article.

Based on the correspondence between these model parameters and the commonly used non-compartmental measures of hormone secretion, we argue that the stress *reactivity* is best reflected by the difference between the individual maxima and minima of cortisol concentrations (*MinMax*), whereas stress *recovery* is best reflected by the minimal concentration (*Cmin*). The secretory delay and the stress-unrelated cortisol fluctuations are best reflected by the time of the concentration maximum (*Tmax*) and the initial concentration (*Cinit*), respectively. When *Tmax* is the non-compartmental parameter of major interest, researchers should consider that its precise assessment requires a higher sampling frequency during the time period at which maximal cortisol concentrations can be expected (according to the conservative estimates of the ODE model, 95% of all peaks are supposed to occur in between 5 min and 32 min after the onset of stress exposure).

Finally, the compiled results call for caution when interpreting the popular *AUC* measures of cortisol secretion because they seem to be comprised of variance due to stress *reactivity* and stress-unrelated secretory activity of the HPA axis. However, this confounding risk can be alleviated by statistical adjustments for *Cinit*. Because the statistical power to detect outcome-associations also relies on such analytical considerations, we hope that the above given recommendations will become valuable for future studies that seek to validly disentangle the specific process components of the cortisol stress response.

# 7. References

Alexander, N., Wankerl, M., Hennig, J., Miller, R., Zänkert, S., Steudte-Schmiedgen, S., Stalder, T., Kirschbaum, C., 2014. DNA methylation profiles within the serotonin transporter gene moderate the association of 5-HTTLPR and cortisol stress reactivity. *Translational Psychiatry*, *4*, e443.

Balodis I.M., Wynne-Edwards, K.E., Olmstead, M.C., 2010. The other side of the curve: examining the relationship between pre-stressor physiological responses and stress reactivity. *Psychoneuroendocrinology, 35*, 1363-1373.

Bonate, P.L., 2011. *Pharmacokinetic-Pharmacodynamic Modeling and Simulation*. Springer, New York.

Box, G.E.P., Draper, N.R., 1987. *Empirical Model-Building and Response Surfaces*. Wiley, Chichester.

Miller et al. (in press). https://doi.org/10.1016/j.psyneuen.2017.12.019

Brown, E.N., Meehan, P.M., Dempster, A.P., 2001. A stochastic differential equation model of diurnal cortisol patterns. *American Journal of Physiology - Endocrinology and Metabolism*, *280*, E450-E461.

Buning, J.W., Touw, D.J., Brummelman, P., Dullaart, R.P., van den Berg, G., van der Klauw, M.M., Kamp, J., Wolffenbuttel, B.H.R., van Beek, A.P., 2017. Pharmacokinetics of oral hydrocortisone – results and implications from a randomized controlled trial. *Metabolism, 71*, 7-16.

Cox, C., Chu, H., Schneider, M.F., Muñoz, A., 2007. Parametric survival analysis and taxonomy of hazard functions for the generalized gamma distribution. *Statistics in Medicine*, *26*, 4352-4374.

Chrousos, G.P., 2009. Stress and disorders of the stress system. *Nature Reviews Endocrinology*, *5*, 374-381.

Czock, D., Keller, F., Rasche, F.M., Häussler, U., 2005. Pharmacokinetics and pharmacodynamics of systemically administered glucocorticoids. *Clinical Pharmacokinetics, 44*, 61-98.

Deboeck, P.R., Bergeman, C.S., 2013. The reservoir model: A differential equation model of psychological regulation. *Psychological Methods*, *18*, 237-256.

Dickerson, S.S., Kemeny, M.E., 2004. Acute stressors and cortisol responses: a theoretical integration and synthesis of laboratory research. *Psychological Bulletin*, *130*, 355-391.

Donnet, S., Samson, A., 2013. A review on estimation of stochastic differential equations for pharmacokinetic/pharmacodynamic models. *Advanced Drug Delivery Reviews*, *65*, 929-939.

Engert, V., Efanov, S.I., Duchesne, A., Vogel, S., Corbo, V., Pruessner, J.C., 2013. Differentiating anticipatory from reactive cortisol responses to psychosocial stress. *Psychoneuroendocrinology*, *38*, 1328-1337.

Engert, V., Vogel, S., Efanov, S.I., Duchesne, A., Corbo, V., Ali, N., Pruessner, J.C., 2011. Investigation into the cross-correlation of salivary cortisol and alpha-amylase responses to psychological stress. *Psychoneuroendocrinology*, *36*, 1294-1302.

Ette, E.I., Onyiah, L.C., 2002. Estimating inestimable standard errors in population pharmacokinetic studies: the bootstrap with Winsorization. *European Journal of Drug Metabolism and Pharmacokinetics*, *27*, 213-224.

Federenko, I.S., Nagamine, M., Hellhammer, D.H., Wadhwa, P.D., Wüst, S., 2004. The heritability of hypothalamus pituitary adrenal axis responses to psychosocial stress is context dependent. The *Journal of Clinical Endocrinology & Metabolism*, *89*, 6244-6250.

Fekedulegn, D.B., Andrew, M.E., Burchfiel, C.M., Violanti, J.M., Hartley, T.A., Charles, L.E., Miller, D.B., 2007. Area under the curve and other summary indicators of repeated waking cortisol measurements. *Psychosomatic Medicine*, *69*, 651-659.

Ferreira-González, I., Permanyer-Miralda, G., Busse, J.W., Bryant, D.M., Montori, V. M., Alonso-Coello, P., Walter, S.D., Guyatt, G.H., 2007. Methodologic discussions for using and interpreting composite endpoints are limited, but still identify major concerns. *Journal of Clinical Epidemiology*, *60*, 651-657.

Gabrielsson, J., Weiner, D., 2006. *Pharmacokinetic and Pharmacodynamic Data Analysis: Concepts and Applications* (4th Ed.). Swedish Pharmaceutical Press, Stockholm.

Gabrielsson, J., Weiner, D., 2012. Non-compartmental analyses, in: Reisfeld, B., Mayeno, A. N. (Eds.), *Computational Toxicology*. Springer, New York, pp. 377-389.

Gatti, R., Antonelli, G., Prearo, M., Spinella, P., Cappellin, E., Elio, F., 2009. Cortisol assays and diagnostic laboratory procedures in human biological fluids. *Clinical Biochemistry*, *42*, 1205-1217.

Gelman, A., Carlin, J. B., Stern, H.S., Rubin, D.B., 2014. *Bayesian Data Analysis* (Vol. 2). Chapman & Hall, Boca Raton, FL.

Gelman, A., Lee, D., & Guo, J., 2015. Stan: A probabilistic programming language for Bayesian inference and optimization. *Journal of Educational and Behavioral Statistics, 40*, 530-543.

Goodman, W.K., Janson, J., Wolf, J.M, 2017. Meta-analytical assessment of the effects of protocol variations on cortisol responses to the Trier Social Stress Test. *Psychoneuroendocrinology*, 80, 26-35.

Gröschl, M., Köhler, H., Topf, H.G., Rupprecht, T., Rauh, M., 2008. Evaluation of saliva collection devices for the analysis of steroids, peptides and therapeutic drugs. *Journal of Pharmaceutical and Biomedical Analysis*, *47*, 478-486.

Gröschl, M., 2008. Current status of salivary hormone analysis. *Clinical Chemistry*, *54*, 1759-1769.

Hankin, B.L., Badanes, L.S., Smolen, A., Young, J.F., 2015. Cortisol reactivity to stress among youth: Stability over time and genetic variants for stress sensitivity. *Journal of Abnormal Psychology*, *124*, 54.

Humphreys, L.G., Montanelli, R. G., 1975. An investigation of the parallel analysis criterion for determining the number of common factors. *Multivariate Behavioral Research, 10*, 193-205.

Johnson, T.D., 2007. Analysis of pulsatile hormone concentration profiles with nonconstant basal concentration: A Bayesian approach. *Biometrics*, *63*, 1207-1217.

Miller et al. (in press). https://doi.org/10.1016/j.psyneuen.2017.12.019

Khoury, J.E., Gonzalez, A., Levitan, R.D., Pruessner, J.C., Chopra, K., Santo Basile, V., Masellis, M., Goodwill, A., Atkinson, L., 2015. Summary cortisol reactivity indicators: Interrelations and meaning. *Neurobiology of Stress*, *2*, 34-43.

Kirschbaum, C., Hellhammer, D.H., 1994. Salivary cortisol in psychoneuroendocrine research: recent developments and applications. *Psychoneuroendocrinology*, *19*, 313-333.

Kirschbaum, C., Pirke, K.M., Hellhammer, D. H., 1993. The 'Trier Social Stress Test' – a tool for investigating psychobiological stress responses in a laboratory setting. *Neuropsychobiology*, *28*, 76-81.

Klim, S., Mortensen, S.B., Kristensen, N.R., Overgaard, R.V., Madsen, H., 2009. Population stochastic modelling (PSM) - an R package for mixed-effects models based on stochastic differential equations. *Computer Methods and Programs in Biomedicine*, *94*, 279-289.

Klim, S., Mortensen, S., 2008. Validation of Population Stochastic Modelling R Package in Comparison with NONMEM VI. http://www2.imm.dtu.dk/projects/psm/PSM_Validation.pdf (accessed 3 October 2017).

Koolhaas, J.M., Bartolomucci, A., Buwalda, B.D., De Boer, S.F., Flügge, G., Korte, S. M., Meerlo, P., Murison, R., Olivier, B., Palanza, P., Richter-Levin, G., Sgoifo, A., Steimer, T., Stiedl, O., van Dijk, G., 2011. Stress revisited: a critical evaluation of the stress concept. *Neuroscience & Biobehavioral Reviews*, *35*, 1291-1301.

Kraan G.P., Dullaart R.P., Pratt, J.J., Wolthers, B.G., de Bruin, R., 1997. Kinetics of intravenously dosed cortisol in four men: concequences for the calculation of the plasma cortisol production rate. *Journal of Steroid Biochemistry & Molecular Biology, 63*, 139-146.

Kudielka, B.M., Hellhammer, D.H., Wüst, S., 2009. Why do we respond so differently? Reviewing determinants of human salivary cortisol responses to challenge. *Psychoneuroendocrinology, 34*, 2-18.

Lentjes, E.G.W.M., Romijn, F.H.T.P.M., 1999. Temperature-dependent cortisol distribution among the blood compartments in man. *The Journal of Clinical Endocrinology & Metabolism*, *84*, 682-687.

Levine, A. Zagoory-Sharon, O., Feldman, R., Lewis, J.G., Weller, A., 2007. Measuring cortisol in human psychobiological studies. *Physiology & Behavior, 90*, 43-53.

Linden, W.L.E.T., Earle, T.L., Gerin, W., Christenfeld, M. (1997). Physiological stress reactivity and recovery: conceptual siblings separated at birth? *Journal of Psychosomatic Research, 42*, 117-135.

Lopez-Duran, N.L., Mayer, S.E., Abelson, J.L., 2014. Modeling neuroendocrine stress reactivity in salivary cortisol: adjusting for peak latency variability. *Stress*, *17*, 285-295.

Miller et al. (in press). https://doi.org/10.1016/j.psyneuen.2017.12.019

Lunn, D.J., Best, N., Thomas, A., Wakefield, J., Spiegelhalter, D., 2002. Bayesian analysis of population PK/PD models: general concepts and software. *Journal of Pharmacokinetics and Pharmacodynamics*, *29*, 271-307.

Miller, R., Plessow, F., Rauh, M., Gröschl, M., Kirschbaum, C., 2012. Comparison of salivary cortisol as measured by different immunoassays and tandem mass spectrometry. *Psychoneuroendocrinology*, *38*, 50-57.

Miller, R., Plessow, F., Kirschbaum, C., Stalder, T., 2013. Classification criteria for distinguishing cortisol responders from nonresponders to psychosocial stress: evaluation of salivary cortisol pulse detection in panel designs. *Psychosomatic Medicine*, *75*, 832-840.

Perogamvros, I., Aarons, L., Miller, A.G., Trainer, P.J., Ray, D.W., 2011. Corticosteroid-binding globulin regulates cortisol pharmacokinetics. *Clinical Endocrinology*, *74*, 30-36.

Pruessner, J.C., Kirschbaum, C., Meinlschmid, G., Hellhammer, D.H., 2003. Two formulas for computation of the area under the curve represent measures of total hormone concentration versus time-dependent change. *Psychoneuroendocrinology*, *28*, 916-931.

R Core Team, 2016. *R: A Language and Environment for Statistical Computing*. R Foundation for Statistical Computing, Vienna.

Savic, R.M., Jonker, D.M., Kerbusch, T., Karlsson, M.O., 2007. Implementation of a transit compartment model for describing drug absorption in pharmacokinetic studies. *Journal of Pharmacokinetics & Pharmacodynamics, 34*, 711-726.

Schlotz, W., Hammerfald, K., Ehlert, U., Gaab, J., 2011. Individual differences in the cortisol response to stress in young healthy men: testing the roles of perceived stress reactivity and threat appraisal using multiphase latent growth curve modeling. *Biological Psychology, 87*, 257-264.

Skrondal, A., Laake, P., 2001. Regression among factor scores. *Psychometrika, 66*, 563-575.

Soetaert, K.E.R., Petzoldt, T., Setzer, R.W., 2010. Solving differential equations in R: package deSolve. *Journal of Statistical Software*, *33*, 1-25.

Spiga, F., Walker, J.J., Terry, J. R., Lightman, S.L., 2014. HPA Axis-Rhythms. *Comprehensive Physiology*, 4, 1273-1298.

Stacy, E.W., 1962. A Generalization of the Gamma Distribution. *Annals of Mathematical Statistics, 33*, 1187-1192.

Stalder, T., Kirschbaum, C., Kudielka, B.M., Adam, E.K., Pruessner, J.C., Wüst, S., Dockray, S., Smyth, N., Evans, P., Hellhammer, D.H., Miller, R., Wetherell, M.A., Lupien, S.J., Clow, A., 2016. Assessment of the cortisol awakening response: expert consensus guidelines. *Psychoneuroendocrinology*, *63*, 414-432.

Miller et al. (in press). https://doi.org/10.1016/j.psyneuen.2017.12.019

Sun, Y.N., Jusko, W.J., 1998. Transit compartments versus gamma distribution function to model signal transduction processes in pharmacodynamics. *Journal of Pharmaceutical Sciences*, *87*, 732-737.

Tunn, S., Möllmann, H., Barth, J., Derendorf, H., Krieg, M., 1992. Simultaneous measurement of cortisol in serum and saliva after different forms of cortisol administration. *Clinical Chemistry*, *38*, 1491-1494.

Urquhart, J., Li, C.C., 1969. Dynamic testing and modeling of adrenocortisol secretory function. *Annals of the New York Academy of Sciences*, *156*, 756-778.

Vandekerckhove, J., Matzke, D., Wagenmakers, E.-J., 2015. Model comparison and the principle of parsimony. In: Townsend, J.T., Busemeyer, J.R. (Eds.) *The Oxford Handbook of Computational and Mathematical Psychology.* doi: 10.1093/oxfordhb/9780199957996.013.1

Voelkle, M.C., Oud, J.H., Davidov, E., Schmidt, P., 2012. An SEM approach to continuous time modeling of panel data: relating authoritarianism and anomia. *Psychological Methods*, *17*, 176.

Wang, Y., 2007. Derivation of various NONMEM estimation methods. *Journal of Pharmacokinetics and Pharmacodynamics*, *34*, 575-593.

Weckesser, L.J., Plessow, F., Pilhatsch, M., Muehlhan, M., Kirschbaum, C., Miller, R., 2014. Do venepuncture procedures induce cortisol responses? A review, study, and synthesis for stress research. *Psychoneuroendocrinology*, *46*, 88-99.

Yafune, A., Ishiguro, M., 1999. Bootstrap approach for constructing confidence intervals for population pharmacokinetic parameters. I: A use of bootstrap standard error. *Statistics in Medicine*, *18*, 581-599.

Xiong, Y., James, D., He, F., Wang, W., 2015. PMXstan: An R Library to Facilitate PKPD Modeling with Stan. *Journal of Pharmacokinetics and Pharmacodynamics, 42*, S11.

## 8. Appendix

### A. The (generalized) Gamma distribution

Cox et al. (2007) report the density function of a generalized Gamma distribution *gGamma*(θ, σ, λ) that transitions into a Weibull distribution when λ = 1, or into a log-normal distribution when λ = 0. After substitution of θ by log(μ) the density function of this generalized Gamma distribution (Eq. A1) can be simplified to the density function of a *Gamma*(μ, σ) distribution (Eq. A2) where λ = σ. As outlined in section 2.3, the average time of cholesterol to be converted in response to stress exposure is thought to correspond to the parameter μ.

Eq. A1

$$f_{gGamma}(t|\mu, \sigma, \lambda) = \frac{|\lambda|}{\Gamma(\lambda^{-2})\sigma t} \left[\frac{1}{\lambda^2}\left(\frac{t}{\mu}\right)^{\lambda/\sigma}\right]^{\lambda^{-2}} e^{-\left(\frac{1}{\lambda^2}\right)\left(\frac{t}{\mu}\right)^{\lambda/\sigma}}$$

Eq. A2

$$f_{Gamma}(t|\mu, \sigma) = \frac{e^{-t/\mu\sigma^2}}{\Gamma(\sigma^{-2})t}\left(\frac{t}{\mu\sigma^2}\right)^{\sigma^{-2}}$$

By contrast, the parameterization of the Gamma distribution referred to in subsection 2.3 is comprised of a shape parameter α = $\sigma^{-2}$ and a rate parameter β = $(\mu\sigma)^{-2}$. The corresponding density function of *Gamma*(α, β) (Eq. A3) forms the algebraic kernel of the product of the transfer rate $k_T$ and the time-dependent, relative amount of cholesterol in the $n^{th}$ transit compartment $R_n(t)$ when α = n+1 and β = $k_T$ (see Eq. 10).

Eq. A3

$$f_{Gamma}(t|\alpha, \beta) = \frac{\beta^\alpha}{\Gamma(\alpha)} t^{\alpha-1} e^{-\beta t}$$

Stacy (1962) reports a similarly parameterized variant of the generalized Gamma distribution *gGamma*(a = 1/β, d = α, p), which will become an ordinary Gamma distribution when p = 1, and a Weibull distribution when d = p.

## B. Model complexity as compared to linear growth curves

A generic property of overly complex models is their capability to account for data features that have only been generated by random noise (i.e. they easily overfit the data). The model complexity of linear growth curves (e.g., polynomials) is simply indicated by the number of model parameters because any data vector can be exhaustively represented as a linear combination of equal length. For the proposed ODE model, however, this simple heuristic is not valid because the nonlinear impact of its parameters constrains the covered state-space based on mechanistic assumptions. Accordingly, complexity of the ODE model (and its corresponding potential to overfit the data) is considerably lower as compared to a linear growth curve comprising the same number of parameters. To assess the complexity of the ODE model, we compared its capability to fit the permuted time series of each individual from the Dresden sample to the fits of different linear growth curves (degree 1 (linear) polynomials – degree 5 (quintic) polynomials). The Figure below shows these model fits (blue boxes) along with their fits to the original time series (white boxes). Notably, the effective complexity of the ODE model was approximately comparable to a linear growth curve model with 3.5 parameters. Moreover, the absolute increase in explained variance of real data was superior to any of the considered linear growth curves.

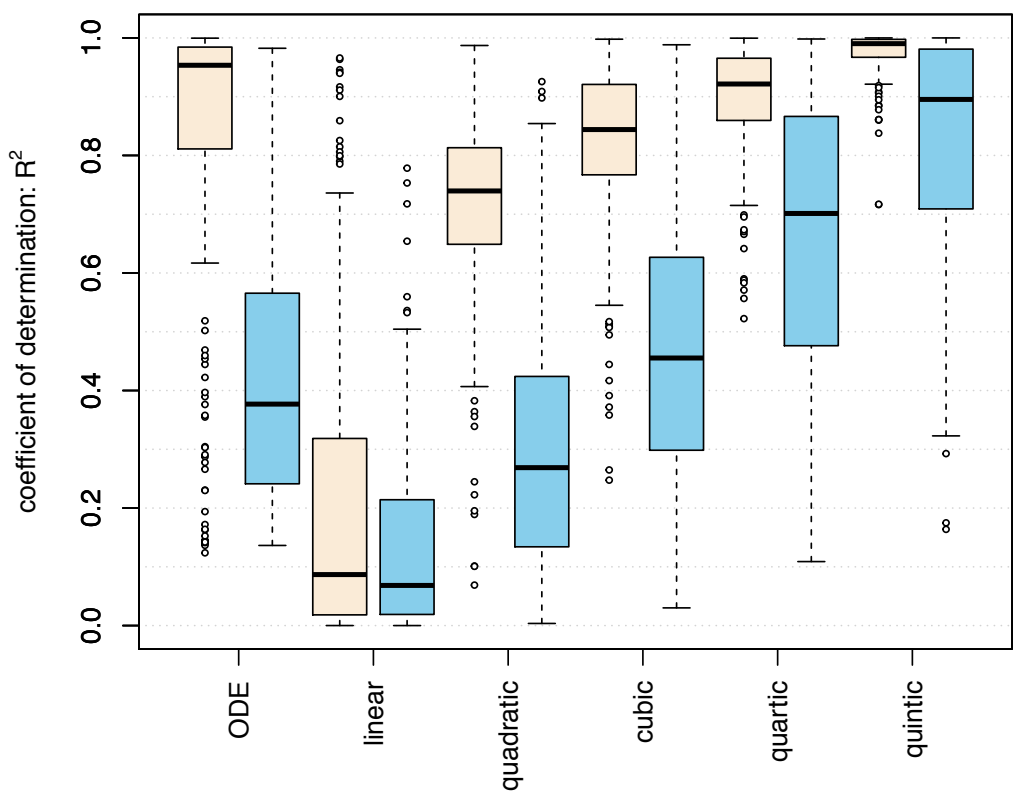

C. Individual cortisol secretion functions *S*(*t*)

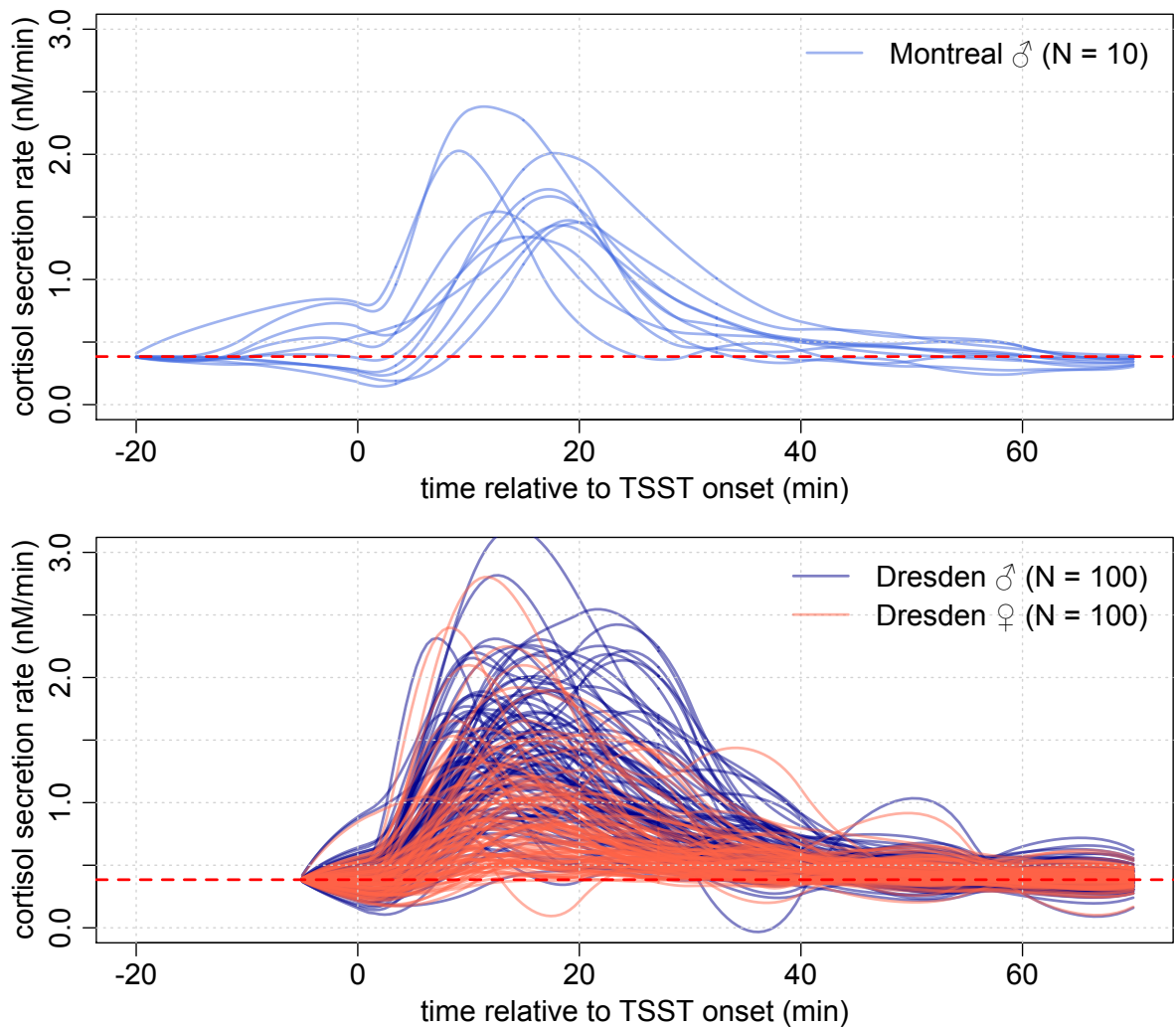

**Figure B1. Cortisol secretion functions from the Montreal sample (upper panel) and the Dresden sample (lower panel) that have been deconvoluted using the SDE population model.**

D. Observed and predicted cortisol concentrations of each participant

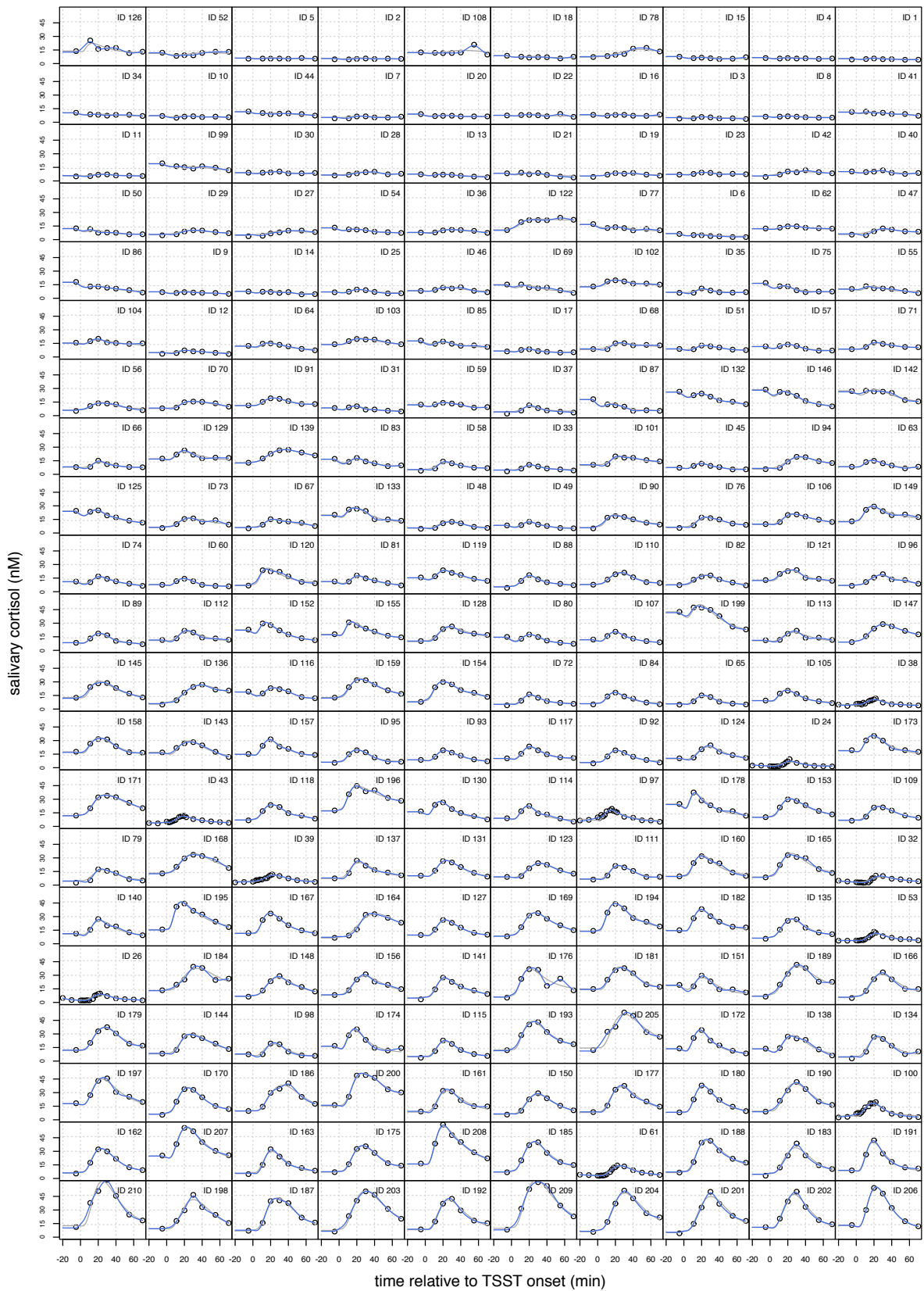

Figure C1. Observed cortisol data and the individual cortisol trajectories as predicted by the ODE population model (grey lines, $R^2$ = 97%) and the SDE population model (blue lines, $R^2$ = 99%). The IDs 24, 26, 32, 38, 39, 43, 53, 61, 97, and 100 denote the 10 male participants from the Montreal sample. The remainder forms the Dresden sample that was composed of 100 males and 100 female participants.